\documentclass[floatfix,reprint,amsmath,amssymb,amsfonts,aps,prd,nofootinbib]{revtex4-1}
\usepackage{dsfont} 
\usepackage{graphicx,bm}
\usepackage{multirow}
\usepackage{color}
\usepackage{enumerate}
\usepackage{placeins}
\usepackage{dcolumn}% Align table columns on decimal point
\usepackage{array}
\usepackage{bbold} %This is needed for double line 1
\usepackage{lipsum} 
\usepackage{slashed}  %for Feynman slash notation
\usepackage[hidelinks]{hyperref}

\newcommand{\be}{\begin{equation}}
\newcommand{\ee}{\end{equation}}
\newcommand{\Tr}{\text{Tr}}

\newcommand{\diag}{\text{diag}}

\newcommand{\vac}{\text{vac}}
\newcommand{\pc}{\text{pc}}

\usepackage{pgf}
\newcommand*{\SectorRadius}{1ex}
\newcommand*{\SectorHalfAngle}{45}
\newcommand*{\SectorLineWidth}{.4pt}
\newcommand*{\sector}{%
  \begin{pgfpicture}
    \pgfpathmoveto{\pgforigin}%
    \pgfpathlineto{\pgfpointpolar{90-(\SectorHalfAngle)}{\SectorRadius}}%
    \pgfarc{90-(\SectorHalfAngle)}{90+\SectorHalfAngle}{\SectorRadius}%
    \pgfpathclose
    \pgfsetlinewidth{\SectorLineWidth}%
    \pgfusepath{stroke}%
  \end{pgfpicture}%
}
\hypersetup{
pdftitle={Sensitivity of finite size effects to the boundary conditions and the vacuum term},
pdfauthor={Gy. Kovács, P. Kovács, P. M. Lo, Gy. Wolf, K. Redlich}
}
\begin{document}

\title{Sensitivity of finite size effects to the boundary conditions and the vacuum term}

\author{Gy\H{o}z\H{o} Kov{\'a}cs}
\email{kovacs.gyozo@wigner.hu}
\affiliation{
Institute for Particle and Nuclear Physics, Wigner Research Centre for Physics, 1121 Budapest, Hungary
}
\affiliation{
Institute of Physics, E\"otv\"os University, 1117 Budapest, Hungary
}

\author{P{\'e}ter Kov{\'a}cs}
\affiliation{
Institute for Particle and Nuclear Physics, Wigner Research Centre for Physics, 1121 Budapest, Hungary
}

\author{Pok Man Lo}
\affiliation{
Institute of Theoretical Physics, University of Wroclaw, \\PL-50204 Wrocław, Poland
}

\author{Krzysztof Redlich}
\affiliation{
Institute of Theoretical Physics, University of Wroclaw, \\PL-50204 Wrocław, Poland
}

\author{György Wolf}
\affiliation{
Institute for Particle and Nuclear Physics, Wigner Research Centre for Physics, 1121 Budapest, Hungary
}

\begin{abstract}
    Finite volume effects are studied both with low-momentum cutoff and with momentum discretization in the framework of an (axial)vector meson extended quark-meson model with Polyakov-loop variables. In the momentum cutoff scenario, the critical endpoint (CEP) moves to lower temperatures and larger quark chemical potentials as the characteristic system size is reduced, however, the treatment of the vacuum term significantly affects its trajectory. The size dependence of the baryon fluctuations is also studied by the kurtosis and the skewness, both of which show moderate dependence on temperature and some dependence on quark chemical potential. The order of the phase transition is also studied near the chiral limit at finite system size and found to be second order only at vanishing explicit breaking. The implementation of the finite size effect with momentum discretization is more complicated and shows peculiar behavior due to the different modes dropping below the Fermi surface and strong dependence on the type of the boundary condition chosen. We found that both the different boundary conditions and the treatment of the vacuum term cause significant changes in the trajectory of the CEP as the characteristic system size is changed.
\end{abstract}

\maketitle

\section{Introduction}

The phase diagram of the strong interaction and, in particular, the existence and location of the critical endpoint (CEP) has been the subject of intense study both in theory and in experiment. However, heavy-ion collisions always suffer from the effect of finite system size, in contrast to field theoretical calculations in the thermodynamic limit. It is expected that a sufficiently small volume can affect the phase diagram and the CEP. 

To study the effects of finite size in theoretical models, it is common to consider the constraints in momentum space imposed by the finite spatial extension. Such constraints can be a discretization or a simple low-momentum cutoff using the phenomenological observation that the lowest modes are the most relevant for the phase transition. In the former case, the momentum integrals are changed to a summation over the modes determined by the boundary conditions, most commonly periodic boundary condition (PBC) and antiperiodic boundary condition (APBC). In the present work, we will focus on these momentum space constraints, but we note that there are other possible implementations of the finite size effects, e.g., the use of different distributions in the thermodynamics \cite{Castano-Yepes:2022nap}.

There have been several attempts in the literature to study the finite volume effects with both discretization and low-momentum cutoff schemes \cite{Palhares:2009tf, Fraga:2010qef, Magdy:2015eda, Magdy:2019frj, Braun:2004yk, Braun:2005gy, Braun:2005fj, Tripolt:2013zfa, Almasi:2016zqf, Klein:2017shl, Bhattacharyya:2012rp, Bhattacharyya:2014uxa, Pan:2016ecs, Wang:2018ovx, Wang:2018qyq, Xu:2019gia, Wan:2020vaj, MataCarrizal:2022gdr, Abreu:2015jya, Abreu:2016ihk, Abreu:2017lxf, Ishikawa:2018yey, Kovacs:2023kcn, Li:2017zny, Luecker:2009bs, Xu:2020loz, Bernhardt:2021iql, Bernhardt:2022mnx}. The treatment of the vacuum term and its finite volume correction varies among the different approaches employed in the analysis.
In linear sigma model (LSM) studies, the vacuum and matter contributions are usually separated after performing the Matsubara sum. However, in the previous works with either discretization \cite{Palhares:2009tf, Fraga:2010qef} or low-momentum cutoff \cite{Magdy:2015eda, Magdy:2019frj}, the vacuum contribution is not present, and therefore its size dependence is not studied. 

However, when the functional renormalization group approach is applied to quark-meson models, there is a renormalized fermionic vacuum integral, which is modified by decreasing size to study the physical quantities at $T=0$ \cite{Braun:2004yk, Braun:2005gy} or the phase transition at zero \cite{Braun:2005fj} and at finite quark chemical potential ($\mu_q$) \cite{Tripolt:2013zfa, Almasi:2016zqf, Klein:2017shl}. 

The Nambu--Jona-Lasinio (NJL) model is nonrenormalizable and therefore requires the specification of an UV regularization scheme to define the model. 
When examining the finite size effects, an additional complexity arises on the infrared side. To address this, the introduction of a low-momentum cutoff can be employed \cite{Bhattacharyya:2012rp, Bhattacharyya:2014uxa, Pan:2016ecs} or via implementing a discretization scheme \cite{Wang:2018ovx, Wang:2018qyq, MataCarrizal:2022gdr}. 

The finite size effects are also studied with the Dyson-Schwinger (DS) equations with a low-momentum cutoff \cite{Li:2017zny} and with discretization \cite{Luecker:2009bs, Xu:2020loz, Bernhardt:2021iql, Bernhardt:2022mnx}. In the DS approach, the modified integrals are more complicated compared to the other discussed models. Because of the different handling of the Matsubara summation, they have renormalized four-momentum integrals.

Combining the results of the different studies, one could conclude that, in general, the CEP is likely to move to lower $T$ and higher $\mu_q$ with decreasing size. However, not only the details of the criticality, e.g., the path and existence of the CEP for very small sizes, but also the size dependence of the phase boundary show different behavior in the different works with different approximations. By implementing several scenarios within a given model, one can investigate whether these differences are due to the approximations used, which will be the goal of the present work.

First, we use a low-momentum cutoff to implement the finite size effects and study the phase transition and thermodynamics at different system sizes. This scenario is also motivated by hadron resonance gas calculations \cite{Karsch:2015zna, Redlich:2016vvb}, where it was found that the volume correction can be correctly reproduced by implementing a low-momentum cutoff in the thermodynamic limit. We will also investigate a less studied problem, the size dependence of baryon fluctuations around the CEP as well, which has been briefly studied in the DS approach \cite{Bernhardt:2021iql}.
To make a comparison between the different implementations and boundary conditions, we also consider the momentum discretization with PBC and APBC, which is widely used, especially in the case of the NJL model and the DS approach. Furthermore, in contrast to other effective models, most previous studies on the quark-meson model have neglected the vacuum contribution to the fermion determinant and hence the size dependence of this term. Therefore we also study how the treatment of the vacuum term---i.e., finite or infinite size---affects the size dependence. 

For this purpose, we use a vector and axial vector meson extended Polyakov quark-meson model (ePQM) with $2+1$ flavors \cite{Kovacs:2016juc}, which is based on \cite{Parganlija:2012fy} with a fermion one-loop contribution, including its---properly renormalized---vacuum and matter part as well. This model gives a good description of the meson phenomenology at $T=0$, $\mu_q=0$, and also of the chiral phase transition at finite temperatures, in good agreement with lattice results, while predicting a CEP and a first-order region for finite $\mu_q$. The ePQM model has been already used for many studies, including the investigation of in-medium vector and axial vector masses \cite{Kovacs:2021kas} and neutron star properties at $T=0$ as the high-density part of a hybrid approach \cite{Takatsy:2019vjs, Kovacs:2021ger}, as well as special limits of the phase diagram such as $N_c\to \infty$ \cite{Kovacs:2022zcl}.

The paper is organized as follows. In Sec.~\ref{Sec:ePQM}, we briefly introduce the ePQM model. Section~\ref{Sec:Cutoff} is dedicated to the low-momentum cutoff scenario of the finite volume effects. After its implementation in the ePQM model, the modification of the phase diagram is discussed. In addition to the importance of the treatment of the vacuum term, we study the size dependence of thermodynamic quantities such as the equation of state (EOS) at $T=0$ and the baryon fluctuations in the vicinity of the CEP. Furthermore, we show the size dependence of the transition and its order in the chiral limit in Appendix~\ref{Sec:chiral_limit}. In Sec.~\ref{Sec:Discretization}, we turn to the discretization scenario. We discuss separately how the size of the vacuum part affects the phase transition and show the size dependence of the CEP with APBC and PBC. Finally, we conclude in Sec.\ref{Sec:Conclusion}.

%%%%%%%%%%%%%%%%%%%%%%%%%%%%%%%%%%%%%%%%%%%%%%%%%%%%%%%%NEWSEC
\section{The extended Polyakov quark-meson model}
\label{Sec:ePQM}

A vector and axial vector meson extended linear sigma model was developed in \cite{Parganlija:2012fy} with $SU(3)_L\times SU(3)_R \times U(1)_V \times U(1)_A$ global symmetry. The model was further extended with constituent quarks and Polyakov-loop variables in \cite{Kovacs:2016juc} (ePQM model), where the main focus was on the thermodynamics and the $T-\mu_B$ phase diagram. 
The Lagrangian of the ePQM model is based on a scalar and a pseudoscalar nonet as well as left- and right-handed vector nonet fields,
\begin{align}
    M =& S+i P=\sum_a \left(S_a + i P_a\right)T_a,  \nonumber\\
    L^\mu =&\sum_a \left(V^\mu_a +A^\mu_a \right)T_a, \;
	R^\mu =\sum_a \left(V^\mu_a -A^\mu_a \right)T_a, \label{Eq:mfields}
\end{align}
where $T_a=\lambda_a/2$ are the generators of the $U(3)$ group, with $\lambda_0=\sqrt{\frac{2}{3}}\mathbb{1}_3$, and $\lambda_i~(i=1,\ldots,8)$ are the Gell-Mann matrices. Following \cite{Kovacs:2016juc}, the Lagrangian can be written as
\begin{widetext}
\begin{align}
    \mathcal{L}_\text{ePQM} =& \Tr \left[ \left( D_\mu M \right)^\dagger \left( D^\mu M \right) \right] - m_0^2 \Tr \left[ M^\dagger M \right] - \lambda_1 \left( \Tr \left[ M^\dagger M \right] \right)^2  - \lambda_2 \Tr \left[ \left( M^\dagger M \right)^2 \right] \nonumber \\
    &+ c \left( \det M + \det M^\dagger \right) + \Tr \left[ H \left( M + M^\dagger \right) \right]  - \frac{1}{4} \Tr \left[ L_{\mu\nu}L^{\mu\nu}+R_{\mu\nu}R^{\mu\nu} \right] \nonumber \\
    &+ \Tr \left[\left(\frac{m_1^2 }{2} +\Delta \right) \left( L_\mu L^\mu + R_\mu R^\mu \right) \right]  +\frac{h_1}{2} \Tr \left[ M^\dagger M \right] \Tr \left[ L_\mu L^\mu +R_\mu R^\mu \right] \label{eq:ELSM_Lag}\\ 
    &+ h_2 \Tr \left[ \left( M R_\mu \right)^\dagger \left( M R^\mu \right) + \left( L_\mu  M \right)^\dagger \left( L^\mu M  \right) \right]  + 2 h_3 \Tr \left[ R_\mu M L^\mu M^\dagger \right] \nonumber \\ 
    &+ i \frac{g_2}{2} \left( \Tr \lbrace L_{\mu\nu} \left[ L^\mu , L^\nu \right] \rbrace + \Tr \lbrace R_{\mu\nu} \left[ R^\mu , R^\nu \right] \rbrace \right)  +\bar\psi \left( i \gamma_\mu \partial^\mu -g_F \mathcal{M}\right) \psi , \nonumber
\end{align}
\end{widetext}
where $\psi=(q_u,q_d,q_s)^T$ are the constituent quarks, and the quark mass matrix is defined as $\mathcal{M}=S+i \gamma_5 P$. The covariant derivatives and the field strength tensors appearing in \eqref{eq:ELSM_Lag} can be written with the help of the left- and right-handed vector fields and the electromagnetic field $A^\mu_e$ as
\begin{align}
    D^\mu =& \partial^\mu M - i  g_1(L^\mu M -M R^\mu) -ie A_e^\mu \left[T_3,M \right], \nonumber\\
    L^{\mu\nu} =& \partial^\mu L^\nu -\partial^\nu L^\mu -ie (A_e^\mu \left[ T_3,L^\nu\right] - A_e^\nu \left[ T_3,L^\mu\right]), \\
    R^{\mu\nu} =& \partial^\mu R^\nu -\partial^\nu R^\mu -ie (A_e^\mu \left[ T_3,R^\nu\right] - A_e^\nu \left[ T_3,R^\mu\right]).\nonumber    
\end{align}

The external fields corresponding to the scalar and vector fields are defined as
\begin{align}
H&=H_0 T_0 + H_8 T_8 = \frac{1}{2} \diag \left( h_{N},h_{N},\sqrt{2}h_{S} \right),\\
\Delta&=\Delta_0 T_0 + \Delta_8 T_8 =  \diag \left( \delta_{N},\delta_{N},\delta_{S} \right) .
\end{align}
Although in nature the isospin symmetry is broken, since its effect is small here we consider the isospin symmetric case (i.e., $m_u=m_d$). As a usual procedure, we assume nonzero vacuum expectation values to scalar fields with zero quantum numbers, that is, $\phi_N=\langle \sigma_N \rangle$ and $\phi_S =\langle \sigma_S \rangle$, which are called the nonstrange and strange condensates. The corresponding scalar fields are subsequently shifted by their expectation values, $\sigma_{N/S}\rightarrow \sigma_{N/S} + \phi_{N/S}$.

To somewhat take into account the effects of confinement, it is usual to introduce Polyakov loops, which can be used to define the Polyakov-loop variables that signal the breaking of the center symmetry (for details, see \cite{Kovacs:2016juc}). Then the grand potential is calculated in a hybrid approximation, where the mesons are treated at tree level and the fermions at one-loop level,
\be \label{eq:grand_pot_def}
\Omega (T,\mu_q )= U(\langle M\rangle ) + \Omega_{\bar qq}^{(0)} (T,\mu_q) + U(\langle \Phi \rangle,\langle \bar \Phi \rangle),
\ee
where the first term is the tree-level mesonic potential, and the second term is the fermionic contribution that includes the effects of the Polyakov loops and consists of a vacuum and a matter part, 
\be
    \Omega_{\bar qq}^{(0)} = \Omega_{\bar qq}^{(0) v} + \Omega_{\bar qq}^{(0) m}
    \label{Eq:grand_pot_quark}
\ee
The third term is the Polyakov-loop potential, for which we use the improved $U = U_{\text{glue}}$, being defined in \cite{Kovacs:2016juc}. Within the current level of mean-field approximation, the exact choice of the Polyakov potential does not change our qualitative results on the phase transition and its finite size dependence as long as the thermodynamics is correctly reproduced. Alternative potentials are available \cite{Fukushima:2003fw, Lo:2013hla, Lo:2021qkw} if one aims at analyzing the fluctuations of the Polyakov loops and matching the curvature masses to lattice QCD data~\cite{Lo:2013hla}.
In the current model, it is necessary to tune the parameter $T_c^\text{glue}=182$ MeV in the $U_{\text{glue}}$ in order to obtain a reasonable chiral pseudocritical temperature at vanishing chemical potential. Even this \textit{ad hoc} procedure can be elevated by taking into account the dressing of the four-quark coupling with quark loops~\cite{Lo:2021buz}, which naturally resolves the problem and provides a natural explanation for the phenomenon of inverse magnetic catalysis. However, the study of fluctuations lies beyond the purpose of this work and we defer such a study for future endeavors.

The model parameters are determined with a $\chi^2$-minimization method using physical quantities like meson masses and decay widths listed in Table V of \cite{Kovacs:2016juc}. The fitted parameters are the bare masses $m_0^2$ and $m_1^2$, the meson-meson couplings $g_1$, $g_2$, $\lambda_1$, $\lambda_2$, $h_1$, $h_2$, and $h_3$, the meson condensates $\phi_{N/S}$, the external field $\delta_{S}$, the $U(1)_A$ anomaly parameter $c$, and the fermion coupling $g_F$. The set of parameters can also be found in \cite{Kovacs:2016juc}.

After parametrization of the model, the thermodynamics is given by the field equations (FEs),
\be
    \frac{\partial \Omega (T,\mu_q )}{\partial \phi_N}=\frac{\partial \Omega (T,\mu_q )}{\partial \phi_S}=\frac{\partial \Omega (T,\mu_q )}{\partial \Phi }=\frac{\partial \Omega (T,\mu_q )}{\partial \bar \Phi}=0 .
\ee 
The explicit form of these in the infinite volume limit is
\begin{widetext}
\begin{align}
    -\frac{d}{d\Phi} \left( \frac{U\left(\Phi, \bar{\Phi}\right)}{T^4} \right) +\frac{6}{T^3} \sum_f \int \frac{d^3p}{(2\pi)^3} \left( \frac{e^{-\beta E_f^-(p)}}{g_f^-(p)} + \frac{e^{-2\beta E_f^+(p)}}{g_f^+(p)} \right) =& 0, \label{Eq:FE_Poly_1}\\
    -\frac{d}{d\bar{\Phi}} \left( \frac{U\left(\Phi, \bar{\Phi}\right)}{T^4} \right) +\frac{6}{T^3} \sum_f \int \frac{d^3p}{(2\pi)^3} \left( \frac{e^{-\beta E_f^+(p)}}{g_f^+(p)} + \frac{e^{-2\beta E_f^-(p)}}{g_f^-(p)} \right) =& 0,\label{Eq:FE_Poly_2}\\
    m_0^2 \phi_N + \left(\lambda_1 + \frac{\lambda_2}{2} \right)\phi_N^3 + \lambda_1\phi_N\phi_S^2 -\frac{c}{\sqrt{2}}\phi_N \phi_S -h_{0N} + \frac{g_F}{2} \sum_{l=u,d} \langle \bar{q_l} q_l\rangle  =& 0, \label{Eq:FE_phiN}\\
    m_0^2 \phi_S + \left(\lambda_1 + \lambda_2 \right)\phi_S^3 + \lambda_1\phi_N^2\phi_S -\frac{\sqrt{2}c}{4}\phi_N^2 -h_{0S} + \frac{g_F}{\sqrt{2}} \langle \bar{q_s} q_s\rangle =& 0, \label{Eq:FE_phiS}
\end{align}
\end{widetext}
where the quark-antiquark condensate is given by
\be \label{eq:qq_cond_full}
\langle \bar q_f q_f\rangle = - 4 N_c m_f \left[\frac{m_f^2}{16\pi^2} \left( \frac{1}{2}+ \log \frac{m_f^2}{M_0^2}\right) + \mathcal{T}_f\right],
\ee 
while the matter part of the tadpole integral reads as
\begin{align} \label{eq:qq_cond_T}
\mathcal{T}_f =- \int \frac{d^3p}{(2\pi)^3} \frac{1}{2E_f(p)} \left( f_f^- (p) + f_f^+ (p) \right) .
\end{align} 
Here
\be 
f_f^\pm (p)  = \frac{\Phi^\pm e^{-\beta E_f^\pm (p)} + 2\Phi^\mp e^{-2\beta E^\pm (p)} e^{-\beta E_f^\pm (p)} }{g_f^\pm (p)}
\ee 
are the modified Fermi-Dirac distribution functions with
\begin{align} \label{eq:def_of_g+-}
  g^+_f =& 1+e^{-3\beta E_f^+} + 3\left[\bar{\Phi} e^{-\beta E_f^+}+\Phi e^{-2\beta E_f^+}\right] .
\end{align}

Solving the field equations above one can study thermodynamics at finite $T$ and/or $\mu_q$ and the phase structure of the strongly interacting matter. To determine the phase boundary between the chirally broken and restored phases, it is common to use the so-called subtracted condensate, which is also calculated on the lattice \cite{Cheng:2007jq} and can be written in our case as
\be \label{eq:subtracted_condensate}
\Delta(T)=\frac{\left( \phi_N - \frac{h_N}{h_S}\phi_S\right)|_{T}}{\left( \phi_N - \frac{h_N}{h_S}\phi_S\right)|_{T=0}}.
\ee 
This quantity can be generalized for finite $\mu_q$ by choosing a proper reference point of the denominator.
The phase boundary is given by the set of inflection points of $\Delta_{\mu_{q,0}}(T)$ [or $\Delta_{T_{0}}(\mu_q)$ in the $\mu_q$ direction].

%%%%%%%%%%%%%%%%%%%%%%%%%%%%%%%%%%%%%%%%%%%%%%%%%%%%%%%%NEWSEC
\section{The ePQM model with low-momentum cutoff}
\label{Sec:Cutoff}

%%%%%%%%%%%%%%%%%%%%%%%%%%%%%%%%%%%%%%%%%%%%%%%%%%%%%%%%newSubSec
\subsection{Implementation of low-momentum cutoff}
\label{sSec:implementetion_low_mom}

 The low-momentum cutoff $\lambda = \pi/L$ can be included by simply introducing a Heavyside function for momentum integrals as
\be \label{eq:lowcutted_integral}
\int\frac{d^3p}{(2\pi)^3} \rightarrow \int \frac{d^3p}{(2\pi)^3} \theta(p-\lambda) .
\ee 
The fermionic vacuum part with the low-momentum cutoff $\lambda$ and the UV regularization $\Lambda$ reads as,
\begin{align} 
\Omega_{\bar qq,\lambda}^{(0)v}=& -2 N_c \sum_{f=u,d,s} \int \frac{d^3p}{(2\pi)^3} E_f (p) \Theta (\Lambda - p) \Theta(p-\lambda) \nonumber\\ =&-\frac{9}{4\pi^2}\Lambda^4 -\frac{3 g^2}{8\pi^2} \left(\phi_N^2 +\phi_S^2 \right) \Lambda^2 
 \label{eq:Omega_vac_ren_lowp}\\
&+ \frac{3g^2}{64\pi^2} \log \left( 2\Lambda e^{-1/4}\right) \left(\phi_N^4+\phi_S^4\right) \nonumber \\
&-\frac{3}{8\pi^2} \sum_{f=u,d,s} \Big[m_f^4 \log \left( \lambda +\sqrt{\lambda^2+m_f^2} \right) \nonumber \\ 
&\qquad\qquad\qquad -\lambda \sqrt{\lambda^2+m_f^2}\left( 2\lambda^2 +m_f^2\right)\Big], \nonumber
\end{align} 
where the $m_f$ tree-level constituent quark masses are
\be 
m_u=m_d=\frac{g_f}{2}\phi_N, \qquad m_s= \frac{g_f}{2}\phi_S.
\ee 
After the same renormalization as in the case of the infinite volume model, the fermionic vacuum contribution reads
\begin{align}
{\Omega_{\bar qq,\lambda,R}^{(0)v}}=&-\frac{3}{8\pi^2} \sum_{f=u,d,s} \Bigg[m_f^4 \log \left( \frac{\lambda +\sqrt{\lambda^2+m_f^2}}{M_0} \right) \nonumber\\
&-\lambda \sqrt{\lambda^2+m_f^2}\left( 2\lambda^2 +m_f^2\right)\Bigg]. \label{eq:ferm_vacuum_lowp}
\end{align}
If $\lambda$ is removed, the resulting expression agrees with the formula in \cite{Kovacs:2016juc}. Using Eq.~\eqref{eq:lowcutted_integral}, the fermionic matter contribution with low-momentum cutoff is given by
\begin{align}
  \Omega_{\bar q q, \lambda}^{(0)\textnormal{T}}(T,\mu_q)  =  -2 T \sum_f\int &
\frac{d^3 p}{(2\pi)^3} \Theta (p-\lambda) \label{eq:fermion_matter_lowcut}\\
&\times \big[\ln g_f^+(p) + \ln g_f^-(p)\big] \text{ ,} \nonumber  
\end{align}
where $ g_f^\pm(p)$ is defined in Eq.~\eqref{eq:def_of_g+-}.

Accordingly, the modified quark-antiquark condensate reads
\begin{align} 
\langle \bar q_f q_f\rangle^{\lambda} &= - 4 N_c m_f \Bigg[ -\frac{1}{8\pi^2} \lambda \sqrt{\lambda^2 +m_f^2}  \label{eq:qq_condensate_vac_lowcut}\\
&+\frac{m_f^2}{16\pi^2} \Bigg( \frac{1}{2}+ \log \frac{(\lambda+ \sqrt{\lambda^2 +m_f^2})^2}{M_0^2}\Bigg) + \mathcal{T}_f^\lambda \Bigg],\nonumber\\
&\equiv \langle \bar q_f q_f\rangle^{\lambda}_{\text{vac}}-4 N_c m_f  \mathcal{T}_f^\lambda \nonumber
\end{align}
with
\begin{align}
\mathcal{T}_f^\lambda &=  -\int \frac{d^3p}{(2\pi)^3} \frac{\Theta(p-\lambda)}{2E_f(p)} \left( f_f^- (p) + f_f^+ (p) \right) \label{eq:tadpole_lowcut}
\end{align}
Now, different linear sizes are chosen, $L\in (2\,\text{fm},\ldots 6 \,\text{fm})$, which corresponds to different low-momentum cutoffs according to $\lambda=\pi/L$. This can be also expressed in MeV by using $1\text{~fm}^{-1}=197.3$~MeV. Then the system of field equations in Eqs.~\eqref{Eq:FE_Poly_1}--\eqref{Eq:FE_phiS} is solved with all the integrals replaced according to Eqs.~\eqref{eq:lowcutted_integral} and \eqref{eq:qq_condensate_vac_lowcut}. For this we also need a parameter set, which is taken from \cite{Kovacs:2016juc}.

The curvature meson masses for the scalars and pseudoscalars---which include one-loop fermion corrections---are also modified in case of low-momentum cutoff. Details can be found in Sec.~B of \cite{Kovacs:2016juc}; here only the modifications are highlighted. The curvature meson masses are defined as
\be
  m^2_{i,{ab}} = \frac{\partial^2 \Omega(T,\mu_q)}{\partial
    \varphi_{i,a} \partial \varphi_{i,b}}
  \bigg|_\textnormal{min}=\mathfrak{m}^2_{i,ab}+\Delta m^2_{i,ab}+\delta
  m^2_{i,ab},
\label{Eq:M2i_ab}
\ee
where the tree-level part $\mathfrak{m}^2_{i,ab}$ is unchanged and can be found in Table I of \cite{Kovacs:2016juc}. In the matter part $\delta m_{ij}^2$, the formulas of Table I of \cite{Kovacs:2016juc} can be used by changing only the tadpole $\mathcal{T}_f$ and bubble $\mathcal{B}_f$ integrals\footnote{We note that in \cite{Kovacs:2016juc} an unconventional sign was used for the matter part of the tadpole integral. Therefore, there is a $-1$ factor difference in the definition of $\mathcal{T}_f$ compared to the cited work.} to their volume-dependent version $\mathcal{T}_f^\lambda$ in Eq.~\eqref{eq:tadpole_lowcut}, and $\mathcal{B}_f^\lambda =-d\mathcal{T}_f^\lambda/(dm_f^2)$.\footnote{This relation between the tadpole and bubble integrals holds for vanishing external momenta, applied in our approximation.} Finally, the fermionic vacuum contribution $\Delta m_{ij}^2$ can be written as 
\begin{align}
\Delta m_{ij}^2 &= \sum_{f,f'=u,d,s} \left[ m_{f,ij}^2 \frac{\partial {\Omega_{\bar qq,\lambda,R}^{(0)v}} }{\partial m_f^2}\right. \label{eq:ferm_vac_contr_meson}\\
&+  \left. m_{f,i}^2 m_{f',j}^2 \frac{\partial^2 {\Omega_{\bar qq,\lambda,R}^{(0)v}} }{\partial m_f^2 \partial m_{f'}^2} \right] ,\nonumber
\end{align} 
where $m_{f,i}^2=\partial m_f^2/\partial \varphi_i$ and $m_{f,ij}^2=\partial^2 m_f^2/\partial \varphi_i\partial \varphi_j$ are shorthand notations for the meson field derivatives of the constituent quark masses and can be found in Table II of \cite{Kovacs:2016juc}. Whereas the derivatives of $\Omega_{\bar qq,\lambda,R}^{(0)v}$ read
\begin{align} \label{eq:Omega_derivativ_dm^2}
    \frac{\partial {\Omega_{\bar qq,\lambda,R}^{(0)v}} }{\partial m_f^2} = & -\frac{3}{16 \pi^2} \Big(4 m_f^2 \log \frac{\sqrt{\lambda^2+m_f^2}+\lambda}{M_0} \nonumber\\
     & \qquad \qquad- 4 \lambda \sqrt{ \lambda^2+ m_f^2 }+ m_f^2 \Big) \ , \\
    \frac{\partial^2 {\Omega_{\bar qq,\lambda,R}^{(0)v}} }{\partial m_f^2 \partial m_{f'}^2} = & - \frac{3}{16 \pi ^2} \Big(4 \log \frac{\sqrt{l^2+m_f^2}+\lambda}{M_0} \nonumber \\ \label{eq:Omega_derivativ_dmdm}
    & \qquad \qquad-\frac{4 \lambda}{\sqrt{\lambda^2+m_f^2}} +3 \Big) \delta_{ff'} \ .
\end{align}

%%%%%%%%%%%%%%%%%%%%%%%%%%%%%%%%%%%%%%%%%%%%%%%%%%%%%%%%newSubSec
\subsection{The phase diagram}

\begin{figure}
    \centering
    \includegraphics[width=0.48\textwidth]{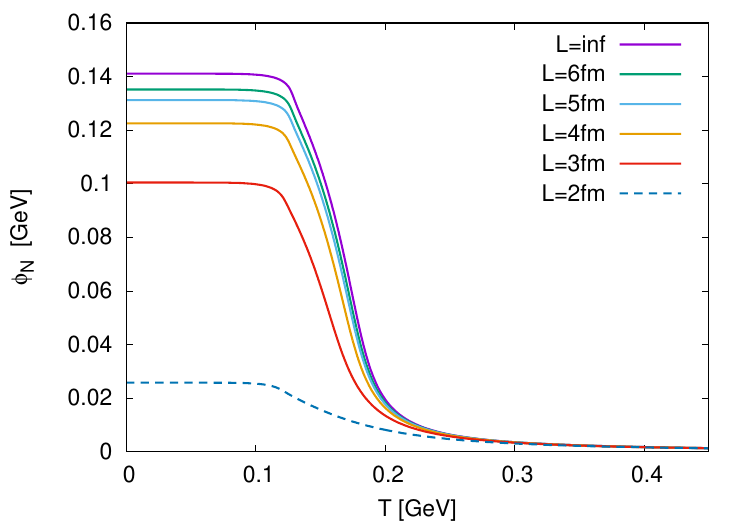}
    \includegraphics[width=0.48\textwidth]{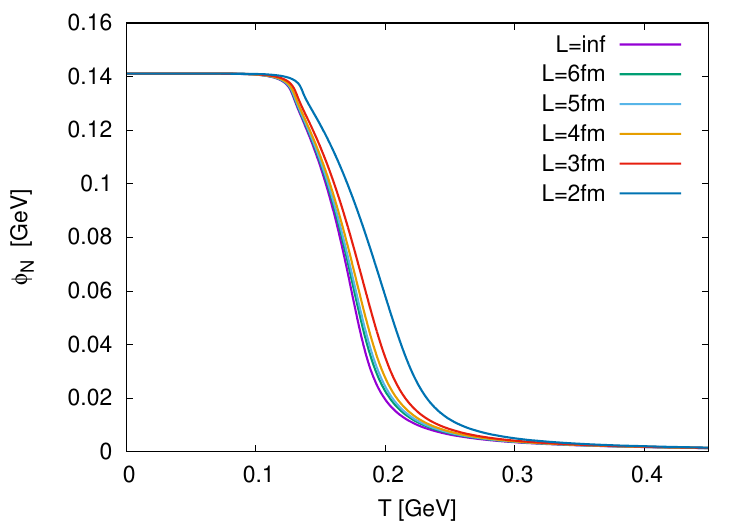}
    \caption{The temperature dependence of $\Phi_N$ for different system sizes $L$ with finite size vacuum term (top) or with infinite size vacuum term (bottom).}
    \label{fig:phiN_by_L_cutoff_both}
\end{figure}

\begin{figure}[hbtp]
    \centering
    \includegraphics[width=0.48\textwidth]{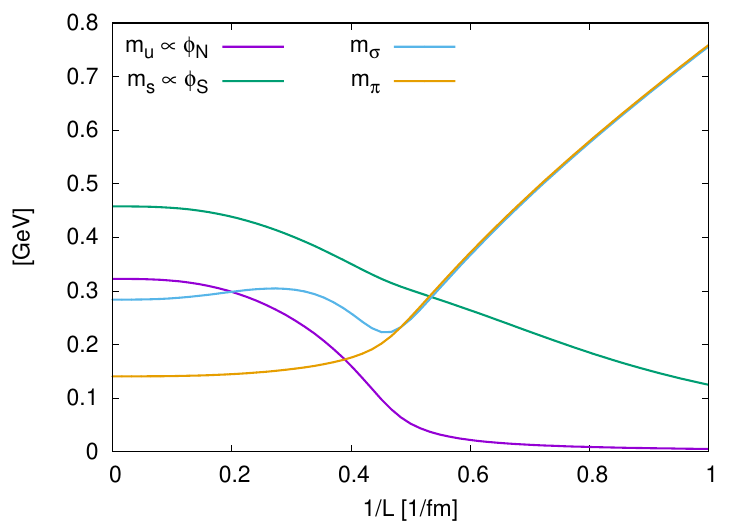}
    \caption{The $1/L$ dependence of tree-level $u$ and $d$ constituent quark masses and one-loop level sigma (or $f_0$) and pion curvature masses in vacuum ($T=\mu_q=0$).}
    \label{fig:hfix_masses_L}
\end{figure}

\begin{figure}[htbp]
    \centering
    \includegraphics[width=0.48\textwidth]{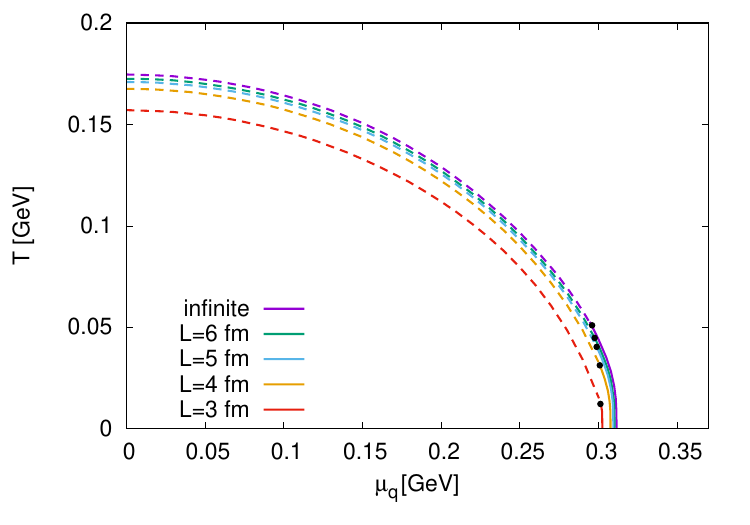}
    \includegraphics[width=0.48\textwidth]{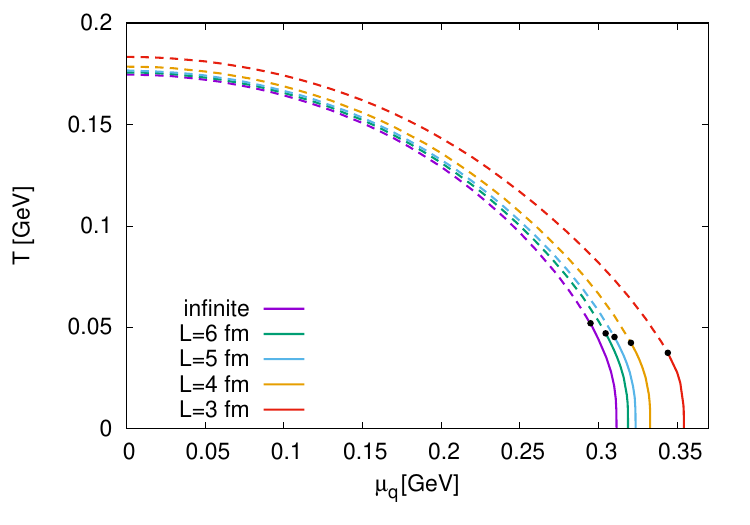}
    \caption{The phase diagram with low-momentum cutoff for different characteristic sizes with finite size  (top) and with infinite size vacuum (bottom). }
    \label{fig:phase_diag_cutoff_both}
\end{figure}

If we change the system size $L$ as described above, the solution of the field equation at $T=\mu_q=0$ gives decreasing values for the meson condensates $\phi_N$ and $\phi_S$ for increasing $L$, which is shown in the top panel of Fig.~\ref{fig:phiN_by_L_cutoff_both} for $\phi_N$. 
It can be seen that the pseudocritical temperature $T_c$ at $\mu_q=0$---defined as the temperature at the inflection point of $\phi_N(T)$ [or $\Delta(T)$]---decreases with decreasing $L$. It is worth noting that on the top figure the curve belonging to $L=2$~fm is already in the region where the pion and sigma masses have already become degenerate (see Fig.~\ref{fig:hfix_masses_L}). It is known that finite size effects are similar to thermal effects if we consider different quantities as a function of $1/L$. In Fig.~\ref{fig:hfix_masses_L} the tree-level masses of the $u$ and $d$ constituent quarks, and the curvature masses of sigma (or $f_0$) and pion are depicted as a function of $1/L$ in the vacuum ($T=\mu_q=0$). The pion and sigma masses become degenerate around $L\approx 2$~fm, which signals chiral symmetry restoration. The transition is of crossover type. We also note that the monotonic increase of the pion mass at small sizes is consistent with results obtained with the Lüscher formalism in chiral perturbation theory \cite{Colangelo:2002hy, Colangelo:2003hf, Colangelo:2005gd}, although in our case the increase turns out to be too large. 
In the top panel of Fig.~\ref{fig:phase_diag_cutoff_both}, the chiral phase boundary is shown for different characteristic sizes. As it was already pointed out, the pseudocritical temperature $T_c$ at $\mu_q=0$ decreases with the decreasing $L$, behavior that has already been seen in the literature \cite{Bhattacharyya:2012rp, Tripolt:2013zfa, Bernhardt:2021iql}; however, opposite behavior was also seen in the case of linear sigma models \cite{Palhares:2009tf, Magdy:2015eda}. This difference comes from the fact that, in our case, the fermionic vacuum contribution is taken into account (see also Sec.~\ref{sSec:vac_at_lowcut}). The second-order CEP moves to smaller temperatures and slightly higher chemical potentials\footnote{Interesting to note that below $L\approx 3$~fm this trend is reversed, just before the CEP disappears} and even disappears around $L\approx 2.5$~fm. This shrinking of the first-order transition line and the absence of the criticality at small volumes is consistent with previous studies \cite{Bhattacharyya:2012rp, Tripolt:2013zfa, Bernhardt:2021iql, Palhares:2009tf, Magdy:2015eda}. When the first-order transition at $T=0$ is present, the critical quark chemical potential $\mu_{q,c}$ also decreases with the decreasing system size; thus the chirally broken region on the $T-\mu_q$ plane becomes smaller with decreasing $L$. 
We note that a similar picture can also be obtained in the case of the chiral limit studied in Appendix~\ref{Sec:chiral_limit}. Instead of a crossover, one finds a second-order phase transition at $\mu_q=0$ ending in a tricritical point, which moves to lower temperatures and chemical potentials with the decreasing system size, similar to the CEP in the case of a nonvanishing explicit symmetry breaking.

%%%%%%%%%%%%%%%%%%%%%%%%%%%%%%%%%%%%%%%%%%%%%%%%%%%%%%%%newSubSec
\subsection{Importance of the vacuum contribution}
\label{sSec:vac_at_lowcut}

\begin{figure}[ht]
    \centering
    \includegraphics[width=0.48\textwidth]{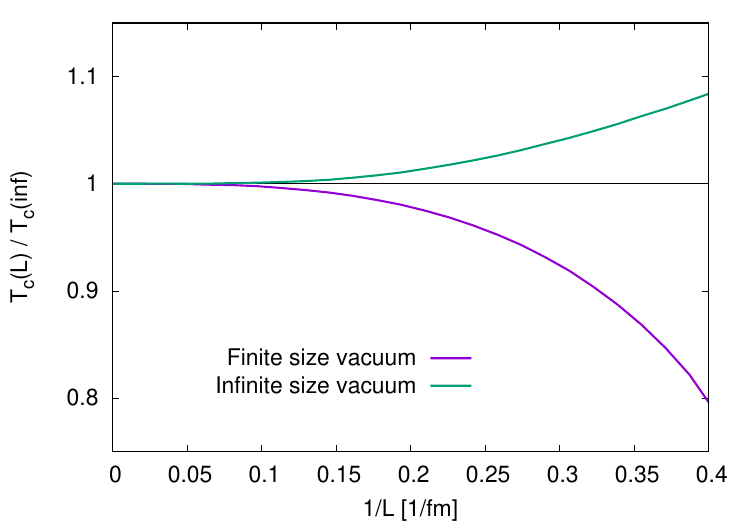}
    \caption{The normalized pseudocritical temperature $T_c/T_{c, \text{inf}}$ for characteristic sizes $L>2.5$ fm with the vacuum contribution treated as finite or infinite.}
    \label{fig:Tpc_comp_lowcut}
\end{figure}

In previous works on finite volume effects within the framework of the linear sigma model, usually the fermionic vacuum contribution was not taken into account (see, e.g., \cite{Palhares:2009tf, Magdy:2015eda, Magdy:2019frj}). To show the importance of the vacuum contribution, we also calculated the $L$ dependence of the condensates and the phase diagram by keeping the vacuum part of the fermion integral at infinite size, i.e., setting $\lambda=0$ in Eqs.~\eqref{eq:ferm_vacuum_lowp} and \eqref{eq:qq_condensate_vac_lowcut}. It is important to note that it is not the mere presence of the vacuum contribution that matters, but the way of its handling,  i.e., its finite or infinite size.\footnote{Already in infinite volume it can be seen that the treatment of the vacuum term, e.g., the regularization used in NJL models, can affect the resulting phase diagram.} In the bottom panel of Fig.~\ref{fig:phiN_by_L_cutoff_both}, the same curves are shown as in the top panel, but for the case where the vacuum is kept at infinite size. It can be seen that the $T=0$ value of $\phi_N$ remains naturally fixed, while the $L$ dependence of the finite temperature integral pushes the pseudocritical temperature to higher temperatures, as can be seen in Fig.~\ref{fig:Tpc_comp_lowcut}, the opposite of our previous case, but consistent with results in \cite{Palhares:2009tf, Magdy:2015eda, Magdy:2019frj}.  
The CEP---shown in the bottom panel of Fig.~\ref{fig:phase_diag_cutoff_both}---moves to smaller temperatures and higher chemical potentials again. However, with no finite size effects in the vacuum contribution, the path of the CEP has a qualitatively different curvature, and consequently, the critical quark chemical potential $\mu_{q,c}$ at $T=0$ increases with the decreasing $L$. We note that in this scenario the CEP does not disappear until $L \approx 0.5$~fm.\footnote{$0.5$~fm is already beyond the reliability of our approach and also smaller than the typical size of physical systems studied in experiments.}

Generally, it can be said that size dependence of the vacuum pushes the system toward the chirally symmetric phase, which is obvious already from the transition in Fig.~\ref{fig:hfix_masses_L}. The broken phase is squeezed down for decreasing size and even disappears completely around $L=2$~fm, if the full fermion one-loop contribution is size dependent. Contrarily, for the vacuum part being fixed at $L=\infty$ the whole phase boundary moves further from the origin in the $T-\mu_q$ plane, i.e., the chirally broken phase even extends. Consequently, at very small sizes the two approaches give qualitatively different results regarding the existence of a chirally broken phase.

%%%%%%%%%%%%%%%%%%%%%%%%%%%%%%%%%%%%%%%%%%%%%%%%%%%%%%%%newSubSec
\subsection{Thermodynamics and baryon fluctuations}

\begin{figure}[htbp]
    \centering
    \includegraphics[width=0.48\textwidth]{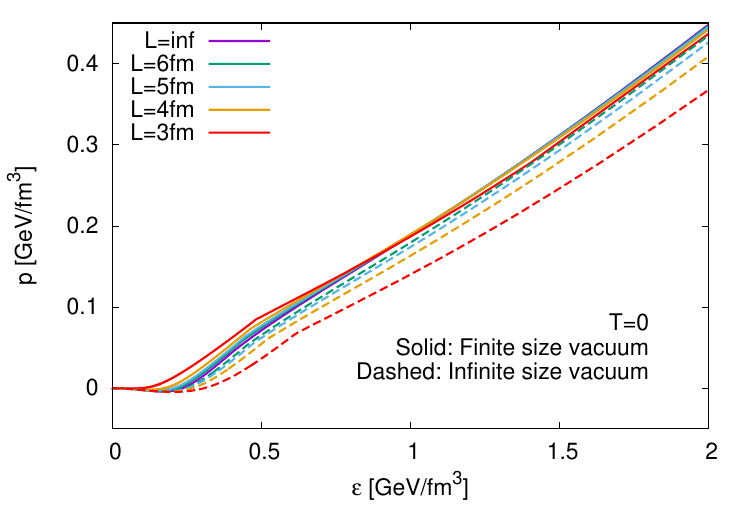}
    \caption{The equation of state, i.e., $p(\varepsilon)$ for infinite and finite volumes of the vacuum term.}
    \label{fig:EoS_cutoff}
\end{figure}

The pressure is defined as 
\be 
p(T,\mu_q)= \Omega(T=0,\mu_q=0) - \Omega(T,\mu_q) ,
\ee 
while other thermodynamic quantities like the entropy density $s=\partial p /\partial T$, the quark number density $\rho_q=\partial p /\partial \mu_q$, and the energy density $\epsilon = -p + Ts + \mu_q \rho_q$ can be derived from it. The equation of state, i.e., the pressure as a function of energy density $p(\epsilon)$ is an important quantity, which is shown at $T=0$ in Fig.~\ref{fig:EoS_cutoff}. It can be seen that the handling of the vacuum contribution affects the EOS. The difference partially comes from the increasing width of the unstable region between the spinodals in $\mu_q$ at $T=0$ and hence from the increasing $\epsilon$ at the transition point for decreasing $L$ in the case of an infinite size vacuum, which pushes downward the $p(\epsilon)$ curve for small sizes.

To study the baryon number fluctuations, the generalized susceptibilities of the baryon number can be defined,
\be 
\chi_n =\left. \frac{\partial^n p/T^4}{\partial (\mu_q /T)^n}\right|_T,
\ee
while the so-called cumulants are related to these susceptibilities as
\be 
C_n = V T^3 \chi_n.
\ee
Since the cumulants are explicitly volume dependent usually their ratios are used, which are equal to the corresponding susceptibility ratios. The most commonly used quantities are the skewness $S\sigma=\chi_3/\chi_2$ and the (excess) kurtosis $\kappa \sigma^2=\chi_4/\chi_2$, where $\sigma$ is the variance, for which $\sigma = \chi_2$ holds. Although explicit volume dependence is canceled in these ratios, they can still have implicit dependence as shown in \cite{Skokov:2012ds, Almasi:2016zqf}, which was already investigated in effective approaches at vanishing chemical potential in \cite{Magdy:2019frj, Bhattacharyya:2014uxa} and with functional methods around the critical endpoint in \cite{Bernhardt:2022mnx}. The cumulants and their ratios are proportional to higher powers of the correlation length, and therefore they could play an important role in the identification of the CEP in experiments \cite{Luo:2015ewa, HADES:2020wpc, PhysRevLett.126.092301}.

\begin{figure}[!htb]
    \centering
    \includegraphics[width=0.48\textwidth]{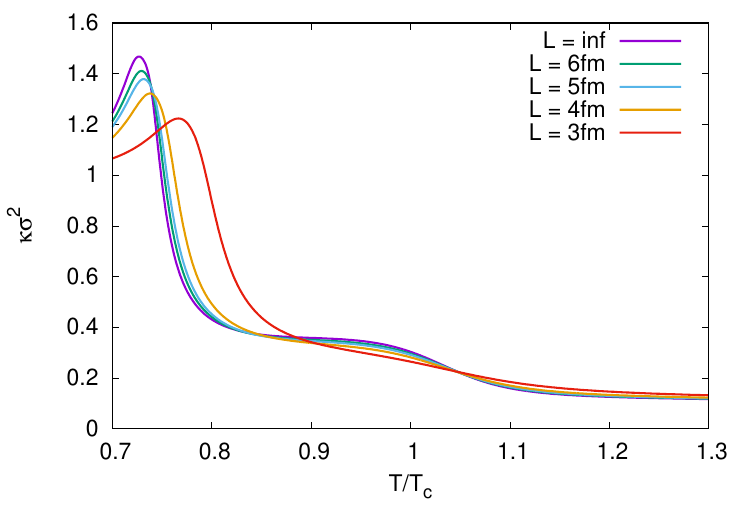}
    \caption{The temperature dependence of the kurtosis $\kappa \sigma^2$ for different system sizes $L$ at $\mu_q=0$.}
    \label{fig:kurt_cutoff_mu0}
\end{figure}

\begin{figure}[tbp]
    \centering\includegraphics[width=0.48\textwidth]{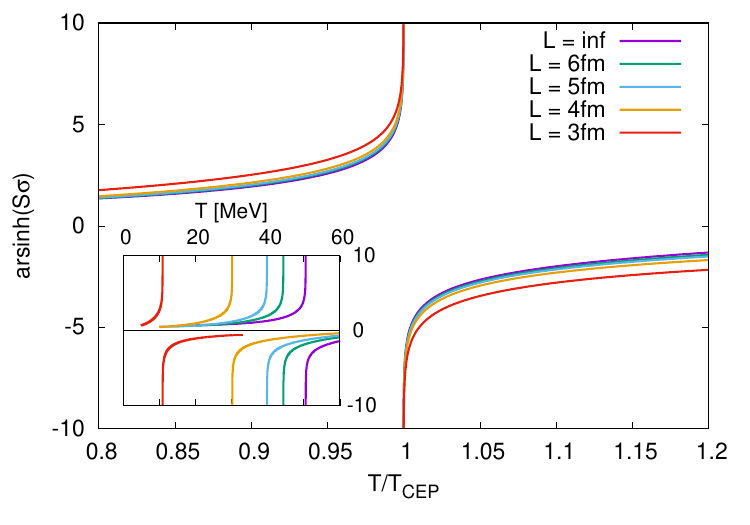}
    \includegraphics[width=0.48\textwidth]{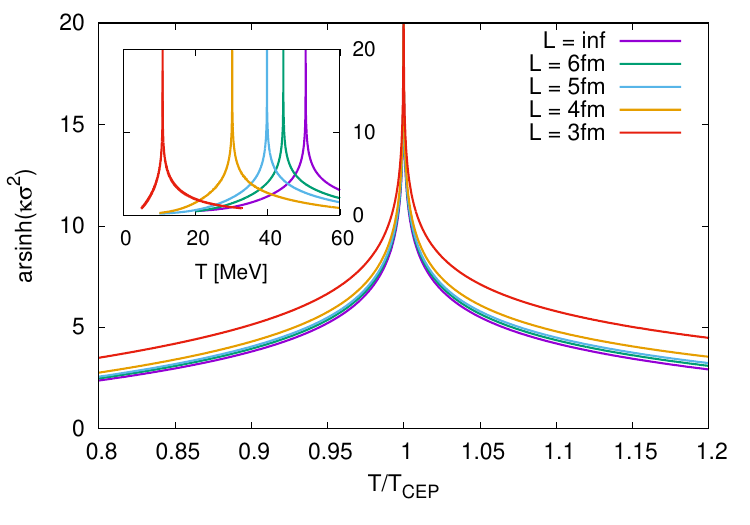}
    \caption{The $T$ dependence of skewness (top) and kurtosis (bottom) for different $L$ values at $\mu_q=\mu_q^\text{CEP}$. The insets show the unscaled $T$ dependence}
    \label{fig:kurt_cutoff_muCEP}
\end{figure}

\begin{figure}[tbp]
    \centering\includegraphics[width=0.48\textwidth]{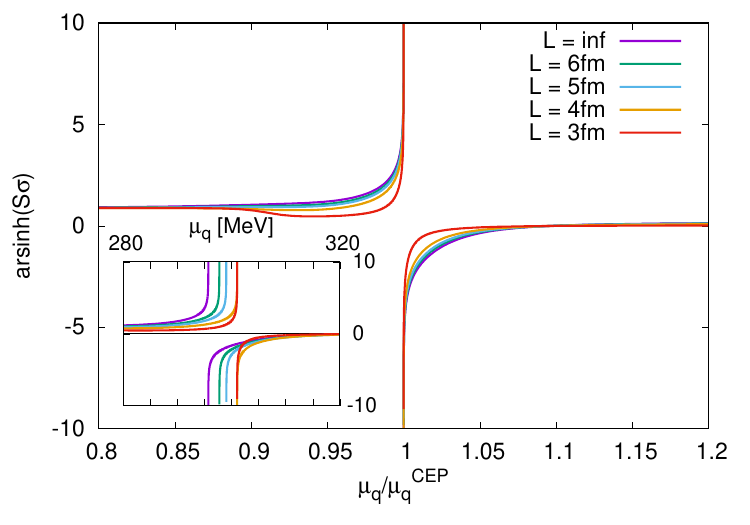}
    \includegraphics[width=0.48\textwidth]{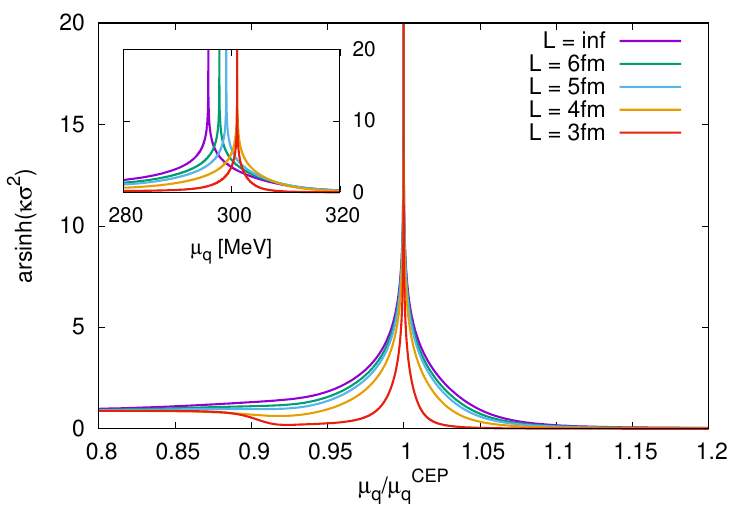}
    \caption{The $\mu_q$ dependence of the skewness (top) and the kurtosis (bottom) at several volume sizes through the critical endpoint. The insets show the unscaled temperature dependence of these quantities. }
    \label{fig:kurt_cutoff_TCEP}
\end{figure}

Baryon fluctuations for infinite volume were already studied in the ePQM model in \cite{Kovacs:2020ihe}. When system size is decreased, the kurtosis at $\mu_q=0$ smoothens as shown in Fig.~\ref{fig:kurt_cutoff_mu0}, which is consistent with previous results of other effective model calculations \cite{Magdy:2019frj, Bhattacharyya:2014uxa}. 
This behavior is in line with the behavior of the $\phi_N$ condensate that also smoothens (see Fig.~\ref{fig:phiN_by_L_cutoff_both}). The temperature dependence of the skewness and the kurtosis is also calculated at $\mu_q=\mu_q^\text{CEP}$, which can be seen in Fig.~\ref{fig:kurt_cutoff_muCEP}. Similarly,  the quark chemical potential dependence of the same quantities at $T=T^\text{CEP}$ are shown in Fig.~\ref{fig:kurt_cutoff_TCEP}.  
It should be noted that the $\mathrm{arsinh}$ function is used to squeeze the function in the vertical direction for better visibility and to be directly comparable to results in the literature \cite{Bernhardt:2022mnx}. It can be seen in Fig.~\ref{fig:kurt_cutoff_muCEP} that there is a slight increase in the kurtosis for temperatures away from $T_{\text{CEP}}$, which becomes larger with decreasing volume. It should be noted that this increase is somewhat artificial in the sense that the $T_\text{CEP}$ values used for the rescale change substantially with $L$. For instance, for $L=3$~fm $T^\text{CEP} \approx 10$~MeV, which is $\approx T^\text{CEP}_\infty/5$. This can be also seen in the inset, where a slight decrease can be observed for smaller and smaller $L$. On the other hand, in Fig.~\ref{fig:kurt_cutoff_TCEP}, in the case of $\mu_q$ dependence, both the rescaled curves and the inset show a decrease away from the transition, which is due to the fact that the change in $\mu_{q,\text{CEP}}$ is moderate with decreasing $L$. 

In conclusion, kurtosis shows a slight change for finite volumes but this effect is most probably originated from the CEP movement toward the chemical potential axis. The increase of $\kappa\sigma^2$ is not remarkable until the CEP comes very close to the chemical potential axis. It is important to note that our result does not contradict what was found in a Dyson-Schwinger approach in \cite{Bernhardt:2022mnx}.\footnote{Beacuse of numerical calculations it is much harder to see the divergence in the DS approach, which gives rise to an artificial depression in the kurtosis close to $T_\text{CEP}$ \cite{Bernhardt:2022mnx}.} There the authors concluded that the kurtosis can be size independent around the CEP, since only a small deviation was found and it had not followed a clear trend in the system size. We also note that their CEP is located at a lower $T_{\text{CEP}}/\mu_{q,\text{CEP}}$ and significantly further away from the chemical potential axis. Note that the temperature dependence of the kurtosis near the critical endpoint has also been studied in a Polyakov--Nambu--Jona-Lasignio (PNJL) model with momentum discretization \cite{Xu:2019gia}. However, due to the unusual phase transition and the multiple CEPs arising in finite volumes when the momentum is discretized (see Sec.~\ref{sSec:Discretization_finite_temp_and_chempot} below), these results cannot be directly compared.

%%%%%%%%%%%%%%%%%%%%%%%%%%%%%%%%%%%%%%%%%%%%%%%%%%%%%%%%NEWSEC
\section{The ePQM model with momentum discretization}
\label{Sec:Discretization}

Fourier transformation naturally implies discretization in momentum space if the system has finite spatial size. The modes that have to be taken into account are determined by the boundary condition imposed on the surface of the system's volume, which therefore also set the lowest momentum mode. In most cases, the finite system is assumed to be a cube with side length $L$ and with periodic or antiperiodic boundary condition. In these cases the momentum integral is substituted with a sum that runs over the modes $p_i = 2n_i \pi/L = n_i\Delta p$ or $p_i = (2n_i+1)\pi/L = (n_i+1/2)\Delta p$, respectively, with $n_i \in \mathds{Z}, i \in (x,y,z)$, while $\Delta p\equiv 2 \pi /L$ is the size of the momentum grid. 

%%%%%%%%%%%%%%%%%%%%%%%%%%%%%%%%%%%%%%%%%%%%%%%%%%%%%%%%newSubSec
\subsection{Implementation of momentum discretization}
\label{sSec:implementation_mom_discr}

In the present work, we implement the periodic boundary condition, the antiperiodic boundary condition, and the periodic boundary condition without the zero mode (PBC-0). It should be noted that other prescriptions are also possible for both the boundary condition (e.g., Dirichlet) and shape of the system (spherical or spheroid, which would also require different BCs) that would give a better description of a real fireball produced in heavy-ion collisions. However, our purpose is to investigate the differences between the results of the different approaches used in the literature, for which the most commonly used PBC and APBC are the best choices.

\begin{figure}[ht]
    \centering
    \includegraphics[width=0.48\textwidth]{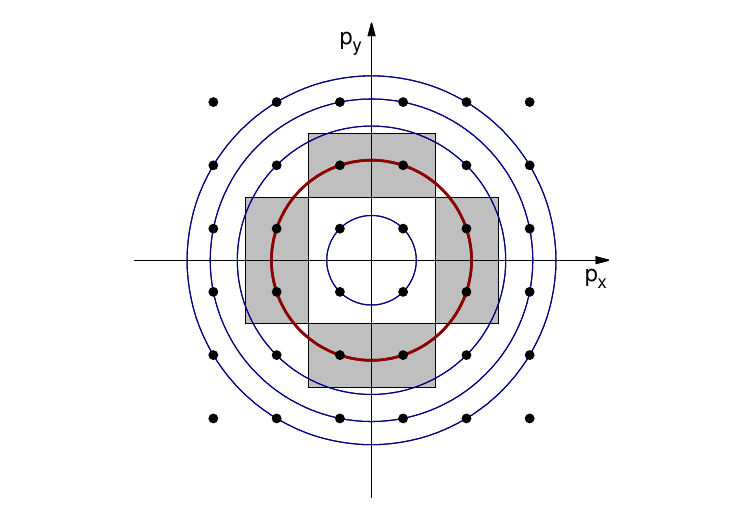}
    \caption{Sketch of the rearrangement of summation for APBC in two dimensions.}
    \label{fig:UV_improvement}
\end{figure}

Summation over the momentum modes can be numerically very expensive for large volumes. To reduce the computational costs, following \cite{ Goecke:2008zh, Bernhardt:2021iql}, the summation is rearranged as follows:
\be \label{eq:sum_cubetosphere}
\sum_{n_x,n_y,n_z=-\infty}^{\infty} K(p_i) = \sum_{j=1}^{\infty} \sum_m K(p_{j},m)
\ee 
for any integral kernel $K(p)$. Here the original grid points of the cube are rearranged in spheres and $p_j$'s are the radii of larger and larger spherical shells in the momentum space as $j$ increases, while $m$ is indexing the grid points on a given shell. This is illustrated in two dimensions for APBC in Fig.~\ref{fig:UV_improvement}, where it can be seen that, for example, for $j=2$ one has $p_2=\sqrt{1/4+9/4}\Delta p$ and a multiplicity of 8 from the trivial sum over $m$. 
It can also be seen in the case of the two-dimensional sketch already, that the distance $\delta p_j = p_j-p_{j-1},~j\geq1$ between two consecutive spheres shows a decreasing trend. Moreover, if the momentum is large enough, an approximate $\delta p_j\propto 1/p_j$ behavior can be found, which allows an additional approximation.
For integration in spherical coordinates, the integrand goes to zero in the matter part and diverges as $p^3$ in the vacuum part of the fermion integral for $p\to \infty$. Therefore, in the matter part, the so-called UV improvement \cite{Xu:2020loz, Bernhardt:2021iql} can be used. In the UV improvement, an intermediate momentum scale $\lambda_\Sigma$ is introduced and the summation is changed to integration for momenta above this scale. This means that for $p<\lambda_\Sigma$ there is a  summation over the discrete modes---ordered now in spheres---while for $p\geq\lambda_\Sigma$ there is an integration with a low-momentum cutoff, similar to our previous scenario. It is worth clarifying the fitting of the boundary of integration to the boundary of summation. Since summation goes over for cubic cells of size $\Delta p^3$ and the integration starts from a sphere, there are some parts of cubes around the boundary that penetrate into the integration region and some other parts that are missing---both for PBC and APBC (see, e.g., Fig.~\ref{fig:UV_improvement} in case of 2D). Thus, there is no common boundary between the summation and the integration. However, for the matter part, the error made by this mismatch is suppressed by the $\delta p_j\propto 1/p_j$ factor and the convergent behavior of the integrand for large momentum. $\lambda_\Sigma$ has to be set for each volume separately, by finding a region where the results are already insensitive to the change of the value of this cutoff. Therefore, the continuum integration can be changed for finite volumes to
\be \label{eq:UVimproved_sum}
\int\frac{d^3p}{(2\pi)^3} \rightarrow \frac{1}{L^3}\sum_{j=1}^{j_\text{max}}\sum_m + \int \frac{d^3p}{(2\pi)^3} \theta(p-\lambda_\Sigma),
\ee 
where $j_\text{max} = \max\{j|p_{j} \leq \lambda_{\Sigma}\}$. Consequently, the fermionic matter contribution can be written as
\begin{align}
  \Omega_{\bar q q, \Sigma}^{(0)\textnormal{T}}(T,\mu_q)  =&  -2 \frac{T}{L^3} \sum_f\sum_{j=1}^{j_\text{max}}\sum_m\big[\ln g_f^+(p_j) + \ln g_f^-(p_j)\big] \nonumber \\
& + \Omega_{\bar q q, \lambda_\Sigma}^{(0)\textnormal{T}}(T,\mu_q). \label{eq:fermion_matter_sum}
\end{align}

\begin{figure}[ht]
    \centering
    \includegraphics[width=0.48\textwidth]{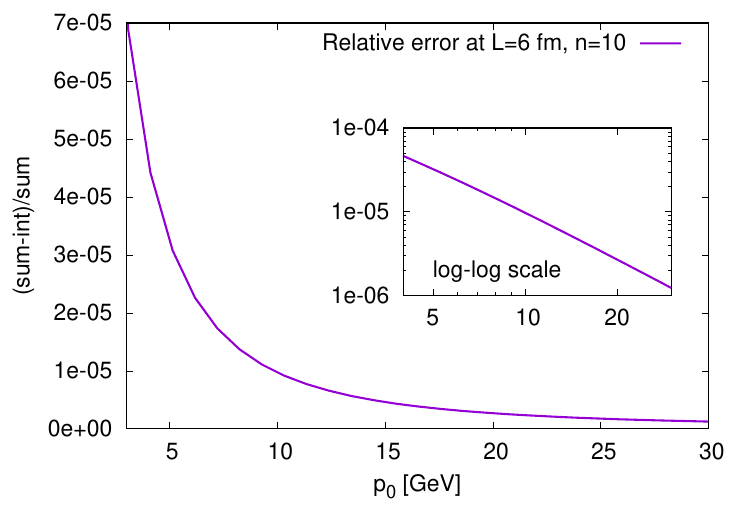}
    \caption{The error made by replacing the summation with integration for the kernel $E(p)$ with $m=300$ MeV. Its decreasing trend allows the use of an UV improvement.}
    \label{fig:Error_cubicUV}
\end{figure}

Unfortunately, in the vacuum part, the integrand is divergent for $p\to \infty$ strong enough to overcome the $\delta p_j\propto 1/p_j$ behavior. Consequently, the error made by changing from a cubic-based summation to a spherical integration at $\lambda_\Sigma$ increases with $\lambda_\Sigma$. Note that this error purely comes from the mismatch of the boundaries of the summation and the integration. This can be seen for APBC as follows: let us keep the original summation starting from some large $p_0$ value up to $p_0 + n \Delta p$, i.e., $p_0 \leq p_i \leq p_0+n \Delta p$, $i \in (x,y,z)$. This corresponds to an integration in Descartes coordinates for the interval $p_0 - \Delta p/2 < p_i < p_0+ (n+1/2)\Delta p$, $i \in (x,y,z)$. The summation and the integration are for exactly the same volume, thus the summation is nothing else but a Riemann sum of the integral. For arbitrary $n$ and $p_0 \gg \Delta p$ it can be shown that with increasing $p_0$ the error made by the summation with midpoint rule compared to the integration decreases, as can be seen in Fig.~\ref{fig:Error_cubicUV}.
In this case, $K(p)=E(p)$ and $E(p) \propto p$ for large $p$. Consequently, considering a well-defined cubic-based boundary between the summation and the integration at a momentum $p_0$, the error, made by changing from summation to integration, can be made arbitrarily small by setting $p_0$ large enough. Therefore, integration can be used in the large momentum part of the fermionic vacuum contribution too. In addition to making the calculation numerically less expensive, this also enables the renormalization of the vacuum part, similar to the infinite volume case. 
The simplest way to use the UV improvement is to change from summation to integration at $\lambda_\Sigma$ in each direction; that is, to sum up to a cube of side length $2\lambda_\Sigma$. Note that the value of $\lambda_\Sigma$ has to be chosen to be half-integer or an integer multiple of the cell size, i.e., $\lambda_\Sigma = (n+1/2) \Delta p$ for PBC and $\lambda_\Sigma = (n+1) \Delta p$ for APBC. To perform the integration that starts from the cube (of side length $2\lambda_\Sigma$), the integration is divided into two parts, the first one from the cube to a sphere of radius $\sqrt{3}\lambda_{\Sigma}$, i.e., the smallest sphere that includes the cube, and the second one which is a usual spherical integral that starts from $p\equiv|p|=\sqrt{3}\lambda_{\Sigma}$. Since the integrand has rotational invariance, in the first part of the integral in spherical coordinates the solid angle integral can be calculated explicitly giving a $p$-dependent solid angle as 
\be 
\Omega^{\lambda_{\Sigma}}(p)=\begin{cases}
\centering
    \Omega^{\lambda_{\Sigma}}_1(p) &\lambda_{\Sigma}<p\leq\sqrt{2}\lambda_{\Sigma}\\
    \Omega^{\lambda_{\Sigma}}_2(p) &\sqrt{2}\lambda_{\Sigma}<p\leq\sqrt{3}\lambda_{\Sigma}\\
\end{cases}
\label{Eq:solid_angle_int}
\ee 
with
\be
\Omega^{\lambda_{\Sigma}}_1(p)=12\pi\displaystyle\frac{p-\lambda_{\Sigma}}{p} \label{Eq:solid_angle_int_p1}
\ee 
and
\begin{align}
    \Omega^{\lambda_{\Sigma}}_2(p)=
    \frac{4}{p}\Bigg( &12p\arccos{\sqrt{\frac{2\lambda_{\Sigma}^2-p^2}{2\lambda_{\Sigma}^2-2p^2}}}
    \nonumber \\
    &+12\lambda_{\Sigma}\arctan\sqrt{p^2/\lambda_{\Sigma}^2-2} \nonumber \\
    &-3\pi(\lambda_{\Sigma}-p)\Bigg), \label{Eq:solid_angle_int_p2}
\end{align}
where there is a change in the expression for the solid angle at $p=\sqrt{2}\lambda_{\Sigma}$, when the sphere touches every middle point of the cube edges. The formula above is derived in Appendix~\ref{Sec:app_solid_angle}. Thus, the fermionic vacuum contribution in Eq.~\eqref{Eq:grand_pot_quark} can be written as three terms, a sum up to $p_i = \pm \lambda_{\Sigma}$, an integration from the cube to the sphere with radius $\sqrt{3}\lambda_\Sigma$, and a renormalized vacuum integral of the form of Eq.~\eqref{eq:ferm_vacuum_lowp} with lower cut-off $\lambda=\sqrt{3}\lambda_\Sigma$,
\begin{align} \label{eq:ferm_vacuum_sum}
    {\Omega_{\bar qq,\lambda_{\Sigma},R}^{(0)v}} =&-2N_c \sum_f \frac{1}{L^3} \sum_{p_x,p_y,p_z}^{|p_i|<\lambda_\Sigma} E_f(p) \nonumber\\
    &-2N_c \sum_f \int_{p_i=\lambda_\Sigma}^{p=\sqrt{3}\lambda_{\Sigma}} \frac{dp }{(2\pi)^3}p^2 \Omega^{\lambda_\Sigma}(p) E_f(p) \nonumber \\
    &+ {\Omega_{\bar qq,\sqrt{3}\lambda_\Sigma,R}^{(0)v}}.
\end{align}

We emphasize that the need for such a complicated adjustment of the intermediate scale for the UV improvement arises from the fact that the discretization does not respect the symmetry of the integration and the renormalization schemes, which has rotational invariance. To overcome this problem, it would be possible to study the finite volume effect with a judicious choice of scheme, so that it is consistent with the symmetry chosen in the renormalization scheme, e.g., with discretization in the radial direction similar to \cite{Lo:2013lca}, where an $O(3)$ scheme was used for the Matsubara sum in $2+1$-dimensional spacetime. This would also be in line with the considerations to implement a rotation invariant, spherical shape of the finite volume system.

\begin{figure}[ht]
    \centering
    \includegraphics[width=0.48\textwidth]{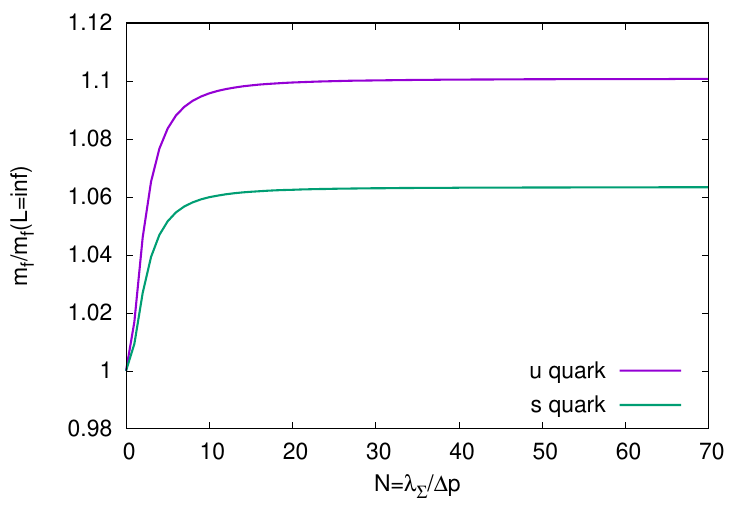}
    \caption{The normalized fermion masses at $L=6$ fm with APBC as a function of the grid size measured by $\lambda_\Sigma/\Delta p$ in the case of UV improvement.}
    \label{fig:test_cubic_spherical}
\end{figure}

To illustrate the stability of the UV improvement with the cubic boundary and the cubic-spherical integration, we show the $\lambda_\Sigma$ dependence of the tree-level constituent quark masses---which are proportional to the meson condensates---at fixed $L$ in Fig.~\ref{fig:test_cubic_spherical}.  
The curves start at 1, since for $\lambda_{\Sigma}=0$ the summation is absent and integration is performed in the whole momentum range, which is equivalent to the infinite size case. In the calculations, we used $\lambda_\Sigma/\Delta p > 50$ to ensure the stability.

The field equations and the curvature meson masses should be modified according to the grand potential in a similar fashion to the low-momentum cutoff case discussed in Sec.~\ref{sSec:implementetion_low_mom}. Every integral derived from the fermionic matter part has to be changed to the UV-improved version given in Eq.~\eqref{eq:UVimproved_sum}. It is straightforward for Eqs.~\eqref{Eq:FE_Poly_1} and \eqref{Eq:FE_Poly_2}, while the matter part of the tadpole integral in Eq~\eqref{eq:qq_cond_T} should be modified as
\be  \label{eq:tadpole_sum}
    \mathcal{T}_f^\Sigma =-\frac{1}{L^3} \sum_{j,m}^{\lambda_\Sigma} \frac{1}{2E_f(p_j)} \left( f_f^- (p_j) + f_f^+ (p_j) \right) + \mathcal{T}_f^{\lambda_\Sigma}
\ee 
with $\mathcal{T}_f^{\lambda_\Sigma}$ being the tadpole with low-momentum cutoff [Eq.~\eqref{eq:tadpole_lowcut}]. Moreover, the fermionic vacuum contribution to the quark-antiquark condensate [first part of Eq.~\eqref{eq:qq_cond_full}] can be derived from Eq.~\eqref{eq:ferm_vacuum_sum}, 
\begin{align} \label{eq:cond_in_FE_sum}
\langle \bar q_f q_f\rangle^\Sigma_\text{vac} =& -2N_c \sum_f m_f\frac{1}{L^3}\sum_{p_x,p_y,p_z}^{|p_i|<\lambda_\Sigma} \frac{1}{E_f(p)}  \nonumber \\
&-2 N_c \sum_f m_f\int_{p_i=\lambda_\Sigma}^{p=\sqrt{3}\lambda_{\Sigma}} \frac{dp }{(2\pi)^3} \frac{p^2\Omega^{\lambda_\Sigma}(p)}{E_f(p)} \nonumber \\ 
&+\langle \bar q_f q_f\rangle^\lambda_\text{vac} \Big|_{\lambda=\sqrt{3}\lambda_\Sigma},
\end{align}
where the last term is the renormalized contribution with a low-momentum cutoff $\lambda=\sqrt{3}\lambda_\Sigma$ defined in Eq.~\eqref{eq:qq_condensate_vac_lowcut}.

The tree-level parts of the meson masses remain unchanged, while for the fermionic matter contribution, $\delta m_{ij}^2$ in Table I of \cite{Kovacs:2016juc} can be used by replacing the tadpole and bubble integrals with $\mathcal{T}_f^\Sigma$ and $\mathcal{B}_f^\Sigma =-d\mathcal{T}_f^\Sigma/(dm_f^2)$, respectively. Moreover, the fermionic vacuum contribution is given similarly as in Eq.~\eqref{eq:ferm_vac_contr_meson}, 
\begin{align}
    \Delta m_{ij}^2 &= \sum_{f,f'=u,d,s} \left[ m_{f,ij}^2 \frac{\partial {\Omega_{\bar qq,\lambda_{\Sigma},R}^{(0)v}} }{\partial m_f^2} \right.\\
     &+ \left.m_{f,i}^2 m_{f',j}^2 \frac{\partial^2 {\Omega_{\bar qq,\lambda_{\Sigma},R}^{(0)v}} }{\partial m_f^2 \partial m_{f'}^2} \right] ,\nonumber
\end{align}
where $m_{f,i}^2$ and $m_{f,ij}^2$ can be found in Table II of \cite{Kovacs:2016juc}. 

%%%%%%%%%%%%%%%%%%%%%%%%%%%%%%%%%%%%%%%%%%%%%%%%%%%%%%%%newSubSec
\subsection{Vacuum contribution}

\begin{figure}[htbp]
    \centering
     \includegraphics[width=0.48\textwidth]{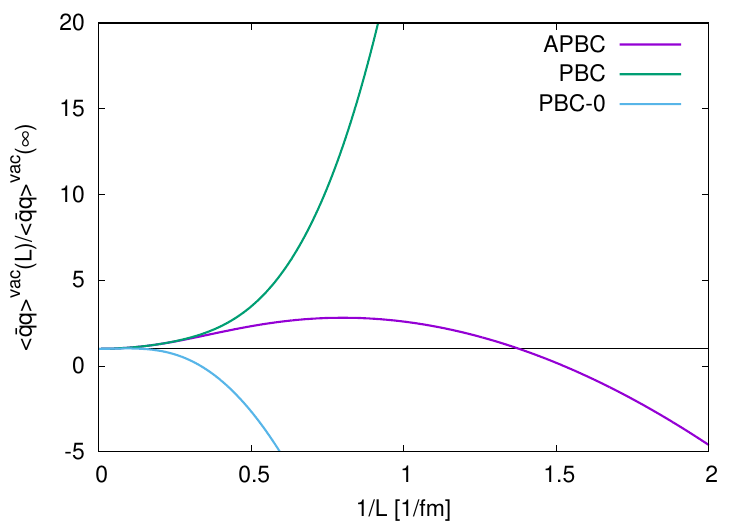}
    \caption{Size dependence of the relative fermion vacuum contribution with different boundary conditions for fixed $m_u=m_u(L=\infty)=322$ MeV.}
    \label{fig:chircondensate_by_L_sums}
\end{figure}

Similar as in Sec.~\ref{sSec:vac_at_lowcut}, we investigated the role of handling of the vacuum contribution, i.e., its infinite or finite size. As discussed in the previous section at $T=\mu_q=0$ the finite size effect to the field equations comes from the discretization of the vacuum part of the chiral condensate $\langle\bar qq\rangle_\vac^\Sigma$ [Eq.~\eqref{eq:ferm_vacuum_sum}]. 
Unfortunately, as the system size is decreased, around $L \lessapprox 4.5$, the solution to the field equations ceases to exist. Contrary to the low-momentum cutoff scenario, here the magnitude of the vacuum contribution to the FEs increases with decreasing $L$, which subsequently causes the disappearance of the common solution to the two FEs [Eqs.~\eqref{Eq:FE_phiN},\eqref{Eq:FE_phiS}] at $T=0$. Note that this kind of instability for large meson condensates is a common feature of the quark-meson models at mean-field level with fermionic vacuum contribution. This is due to the fact that the vacuum term gives a leading contribution to the grand potential, which is $\propto - \bar \sigma^3 \log \bar\sigma$ with the general order parameter $\bar\sigma$, and hence the grand potential is not bounded from below as a function of $\bar\sigma$. When the size decreases in the discretized scenario, the local minimum moves to larger values of the order parameter and collapses with the maximum, hence leaving no extremum of the grand potential. Nevertheless, we can still look at the effect of different boundary conditions as a function of $1/L$ if we fix $m_u$ (and consequently $\phi_N$) and calculate $\langle\bar qq\rangle_\vac^\Sigma(L)$. This can be seen in Fig~\ref{fig:chircondensate_by_L_sums}. 
The most significant contribution to $\langle\bar qq\rangle_\vac^\Sigma$ is at $p=0$, which is taken into account in the case of PBC and omitted in the case of APBC. Consequently, it would be expected that the absolute value of the chiral condensate and thus (via the field equations) $\phi_N$ and $\phi_S$ increases for PBC and decreases for APBC with decreasing $L$. In terms of chiral symmetry, it means that the APBC drives the system to chiral restoration while the PBC increases the spontaneous symmetry breaking. In NJL model studies with discretization of the regularized chiral condensate, such behavior was found in \cite{Wang:2018ovx, Wang:2018qyq, MataCarrizal:2022gdr}. A qualitatively similar result was obtained in a parity doublet model in \cite{Ishikawa:2018yey}. We note that in these cases the vacuum contribution is only regularized with special functions, and no renormalization is performed. However, in the ePQM model for $L>1$ fm, the main difference between the integral at $L=\infty$ and the summation at finite $L$ comes from higher modes, and both boundary conditions give an increasing absolute value for the chiral condensate, as shown in Fig.~\ref{fig:chircondensate_by_L_sums}. Therefore, as $L$ decreases from infinity and the finite size effect appears, the chirally broken phase extends to higher $T$ (and $\mu_q$) not only for PBC but also for APBC. 

It is worth noting that if the vacuum size is kept infinite the solution exists, which will be investigated in the next section. On the other hand, it is still possible to take into account the finite size effect of the vacuum in a restricted way, as it was done in \cite{Palhares:2009tf} for PBC,\footnote{Although in \cite{Palhares:2009tf} no fermionic vacuum part was taken into account at $L=\infty$, its zero mode was added as an extra contribution to the classical potential at finite size.} if we consider only the lowest mode [i.e., $\vec p=(0,0,0)$ for PBC and $\vec p=(1/2,1/2,1/2)$ for APBC]. Practically, this means that only the first term is kept from the sum in Eq.~\eqref{eq:cond_in_FE_sum}. In this way, the solution to the FEs will exist up to $L \approx 2.5$~fm.

Finally, in Fig.~\ref{fig:chircondensate_by_L_sums}, the PBC-0 case is also shown (see, e.g., in \cite{Bernhardt:2021iql}). Here, the low-momentum cutoff caused by the exclusion of the zero mode gives the main modification. Consequently, the absolute value of the chiral condensate decreases as soon as the finite size effect appears around $L=10$ fm, similar to the low-momentum cutoff scenario. In this scenario, the solution to the FEs exists.

%%%%%%%%%%%%%%%%%%%%%%%%%%%%%%%%%%%%%%%%%%%%%%%%%%%%%%%%newSubSec
\subsection{Finite temperature and chemical potential}
\label{sSec:Discretization_finite_temp_and_chempot}

As a consequence of the analysis presented in the previous section, from now on the fermionic vacuum contribution is either kept at infinite volume, or the discretization is taken into account only in the lowest mode, and we implement the summation only for the matter part. To get a better understanding of the size dependence at finite $T$ and $\mu_q$ in addition to the ePQM model, we also considered a simpler linear sigma model presented in \cite{Palhares:2009tf} (called LSM A) and another LSM model with two different parametrizations from \cite{Schaefer:2008hk}. In the latter case, the Lagrangian is similar to Eq.~\eqref{eq:ELSM_Lag} without the (axial) vector meson sector and without fermionic vacuum contribution. The two sets of parameters used here are the one with $m_\sigma=600$ MeV (LSM B) and with $m_\sigma=800$ MeV (LSM C). The parameter sets for LSM B and C can be found in the 11th and 13th row [bottom panel, with $U(1)_A$ breaking] of Table~II in \cite{Schaefer:2008hk}.

Besides the absence of the fermion vacuum contribution in LSM A, LSM B, and LSM C, the location of the CEP is also very different. The ePQM model and LSM C both have a "low-lying" CEP with $\mu_q/T=5-6$ at $(\mu_q, T)=(295,53)$ and $(328,63)$ MeV, respectively, whereas LSM A and B have a ``high-lying'' CEP with $\mu_q/T=2-2.5$ at $(207,100)$ and $(220,92)$ MeV, respectively.\footnote{In baryon chemical potential this means $\mu_B/T=15-18$ for the ``low-lying'' and $\mu_B/T=6-7.5$ for the ``high-lying'' CEPs.} 

\begin{figure}[htbp]
    \centering
     \includegraphics[width=0.48\textwidth]{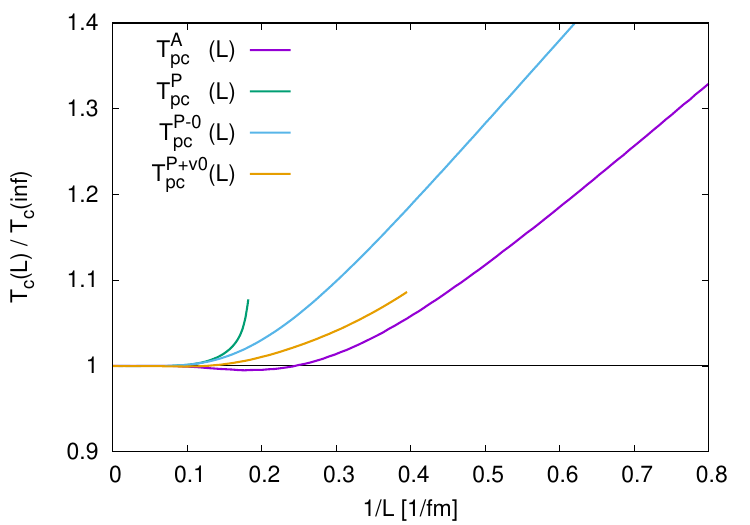}
    \caption{Pseudocritical temperature as a function of $1/L$ for different boundary conditions.} 
    \label{fig:Tc_sum_novac}
\end{figure}

\begin{figure}[tbp]
    \centering
     \includegraphics[width=0.48\textwidth]{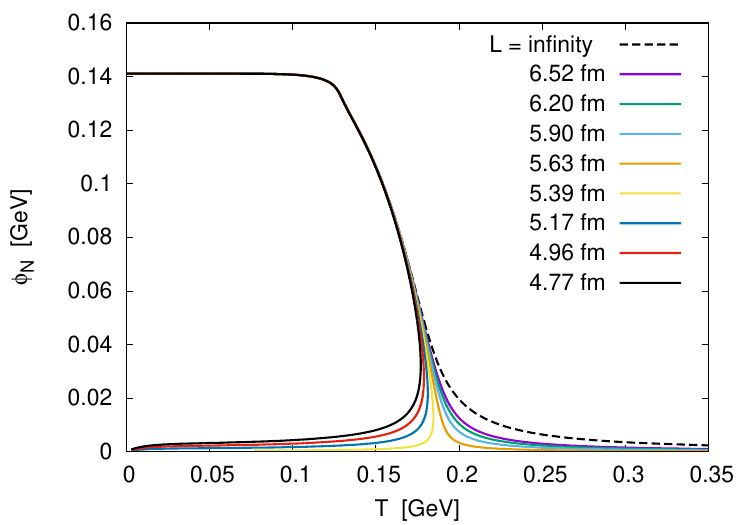}
    \caption{$\phi_N$ versus temperature for different values of $L$ in the case of PBC with vacuum of infinite size.}
    \label{fig:phiN_sumP_nonphys}
\end{figure}

Before turning to the effects of finite size on the CEP in the different model,s we discuss the $\mu_q=0$ and $T \approx 0$ cases in the ePQM model. At $\mu_q=0$ with the vacuum contribution of infinite size, the pseudocritical temperature increases for APBC (after a transient decrease from $1/L=0.1$ to $1/L=0.25 \text{ fm}^{-1}$), PBC and also for PBC-0 as shown in Fig.~\ref{fig:Tc_sum_novac}, labeled as $T_\pc^A$, $T_\pc^P$, and $T_\pc^{P-0}$, respectively.
As can be seen in the case of PBC, the solution ceases to exist at $L\approx 5.46$~fm. This is due to a new unstable solution of the field equations---corresponding to a maximum of the grand potential. The new solution, which is present at low values of $\phi_N$ and $\phi_S$ is caused by the presence of the zero mode of the matter contribution at finite temperatures. Furthermore, the problem with this solution, which is connected to the solution at $T=0$, is that it has no continuation at higher temperatures. This is illustrated in Fig.~\ref{fig:phiN_sumP_nonphys}.
Therefore, the crossover is suddenly replaced by a nonphysical first-order transition. By including the effect of the finite size, at least in the zero mode of the vacuum contribution, the zero mode of the matter part is compensated, and the above problem does not exist. On the other hand, the solvability of the model at $T=0$, $\mu_q=0$ is still limited to $L>2.5$~fm, due to the increasing condensates (caused by the increasing vacuum contribution) discussed in the previous section. The resulting pseudocritical temperature, denoted as $T_\pc^{P+v0}$, is also shown in Fig.~\ref{fig:Tc_sum_novac}. For PBC, contrary to the low-momentum cutoff scenario, the finite size effects on the vacuum contribution (implemented only in the zero mode, but also if the full discretization is applied) enhance the chiral symmetry breaking,\footnote{This means an increase in $\phi_{N/S}$ and thus the restoration happens at larger temperatures} hence the pseudocritical temperature $T_\pc^{P+v0}$ increases with the decreasing system size. 

\begin{figure}[tbp]
    \centering
     \includegraphics[width=0.48\textwidth]{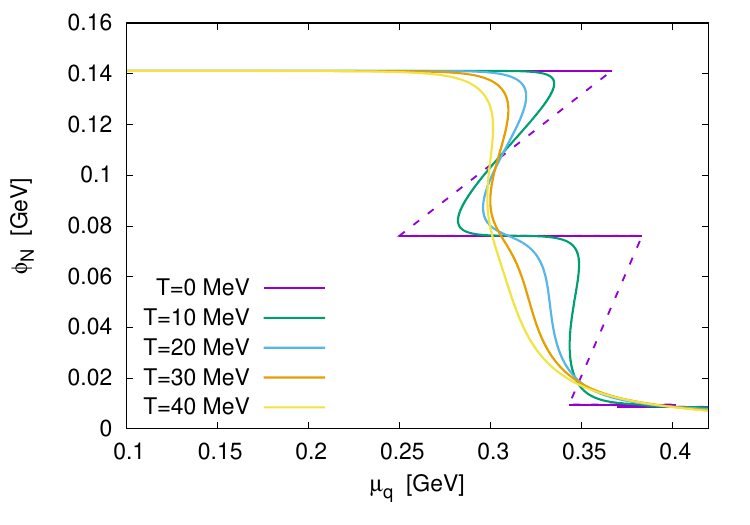}
    \caption{$\phi_N$ as a function of the $\mu_q$ quark chemical potential for different temperatures for APBC.}% $T=0$, $10$, $20$, $30$, and $40$ MeV at $L=6$ fm.} 
    \label{fig:sigma_in_mu_Dp200}
\end{figure}

In the $T \to 0$ limit, the Fermi-Dirac distribution (as a function of $\mu_q$) becomes a step function, which is one below the Fermi surface, i.e., $E(p)<\mu_q$, and zero above. When the momentum space is discretized, only the modes below the Fermi surface contribute. With increasing $\mu_q$, the lowest modes enter below the Fermi surface, causing jumps in the fermionic matter contribution in the FEs and thus leading to drops in the meson condensates.\footnote{This behavior may be an artifact of our assumption, but a similar structure with multistep transitions could be realized, e.g., in some condensed matter systems, as is the case for the Hall conductivity in \cite{Burkov:2011ene}.} Moreover, the decrease of the condensates, and hence of the fermion masses in $E(p)$, enhances the expansion of the Fermi surface (generally driven by the growth of $\mu_q$), which gets closer to or even grows beyond the next modes. This leads to a staircaselike decrease of the condensates with multiple steps, hence multiple inflection points, which look like first-order transitions on top of each other.\footnote{Instead of one phase transition from the chirally broken to the symmetric phase, in the present case there are smaller steps (marked generally by the lines of inflection points, but for a first-order transition also by having multiple solutions at a given $T$ and $\mu_q$). Therefore, the chiral restoration happens in multiple steps, which we will call ``quasitransitions'' since they separate phases with only quantitatively different spontaneous symmetry breaking.} The structure of $\phi_N(\mu_q)$ is shown in Fig.~\ref{fig:sigma_in_mu_Dp200}, 
where $L=6$ fm and APBC is applied for the ePQM model at vanishing and nonzero temperatures. At $T=0$ the horizontal steps are connected by quasisolutions (marked with dashed lines), where $\partial \Omega/\partial \phi_{N/S} = 0$ has a solution, but $\partial \Omega/\partial \phi_{S/N}$ has a discontinuity. These quasisolutions change to real solutions for nonzero temperatures. Since the transition at $T\approx0$ is of first order new second-order points appear on the new quasi transition lines. A similar staircaselike phase transition has also been observed in PNJL model calculations \cite{Xu:2019gia}, where it was named ``quantized first-order phase transition.'' In these models, the gap equation, which is solved self-consistently to find the value of the chiral condensate at a given $T$ and $\mu_q$, formally contains the same integral as the field equations in Eqs.~\eqref{Eq:FE_phiN} and \eqref{Eq:FE_phiS} and thus suffers from the problem of discrete modes entering below the Fermi surface in the same way. For PBC, similar behavior to APBC can be seen when the unphysical first-order transition is not present. From this behavior for both boundary conditions it can be predicted that the low-$T$ and high-$\mu_q$ region of the phase diagram, where the CEP is located, is strongly influenced by the $T\approx0$ structure. The behavior of the CEP is considered separately for APBC and PBC in the following subsections.

We note that we are working in the grand canonical ensemble, where the chemical potential is an external parameter and the Fermi surface is independent of $L$. Meanwhile, the distance between the modes increases with the decreasing size, and hence some of them fall out for a given $\mu_q$. Alternatively, for finite size, one could first formulate the canonical picture with the discretized momentum space and then derive the Fermi surface from the chemical potential defined by the change of the free energy $F$ with the particle number $N$ of the discretized system $\mu_q = dF/dN$. To see whether such a treatment changes the results or not, further investigations are needed.

%%%%%%%%%%%%%%%%%%%%%%%%%%%%%%%%%%%%%%%%%%%%%%%%%%%%%%%%newSubSubSec
\subsubsection{Size dependence of the CEP with APBC}

\begin{figure}[htbp]
    \centering
     \includegraphics[width=0.48\textwidth]{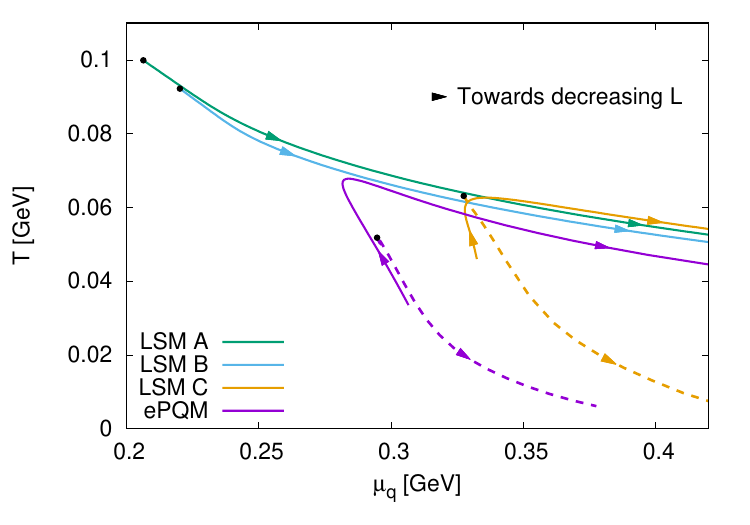}
    \caption{Trajectories of the CEPs on the phase diagram with decreasing $L$ for the LSM A, B, and C and the ePQM models with APBC. Arrows point to lower values of $L$.}
    \label{fig:CEPs_path_APBC}
\end{figure}

\begin{figure}[ht]
    \centering
     \includegraphics[width=0.48\textwidth]{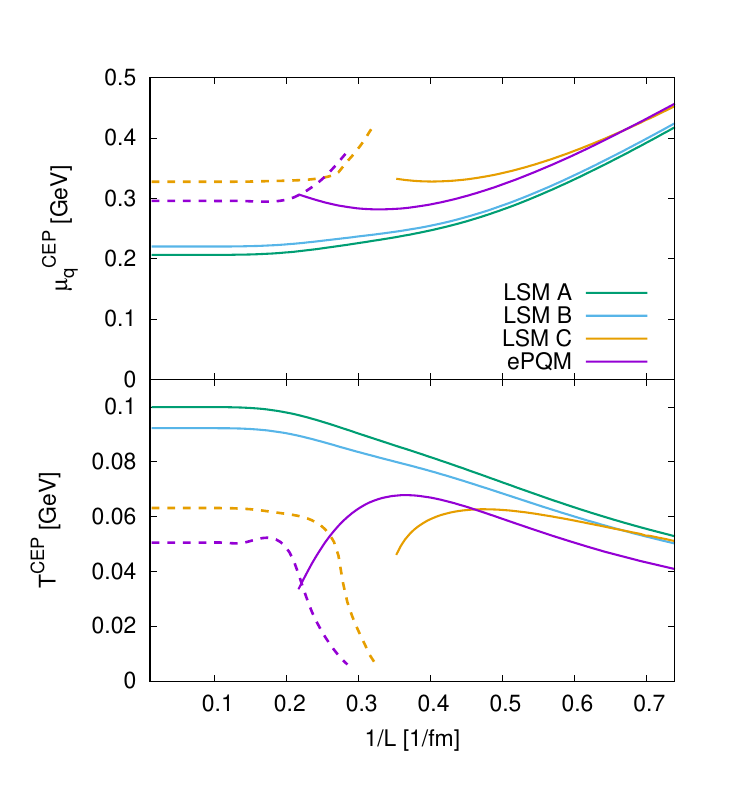}
    \caption{The size dependence of $T^\text{CEP}$ and $\mu_q^\text{CEP}$ for the considered quark-meson models.} 
    \label{fig:TmuCEPs}
\end{figure}

The trajectories of the CEP can be seen in Fig.~\ref{fig:CEPs_path_APBC} as $L$ is decreased in the case of APBC for the models discussed in the previous section.
In each case, the CEP belonging to $L=\infty$ (marked with a black dot) starts to move to lower temperatures and higher quark chemical potentials. This can be seen with the solid line for the LSM A and LSM B models and with the dashed line for the LSM C and ePQM models. However, as discussed in the previous section, the $\phi_N(\mu_q)$ has multiple inflection points (Fig.~\ref{fig:sigma_in_mu_Dp200}). For the ePQM model and LSM C, the inflection point with higher $\mu_q$ and lower $\phi_N$ is connected to the CEP belonging to $L=\infty$ (dashed trajectory), but at some $L$ value this part of the curve becomes a crossover. On the other hand, the second inflection point (with lower $\mu_q$ and higher $\phi_N$), which coexist in some region with the first one, separates the crossover and first-order regions for $L\lessapprox 4.5$~fm; therefore it is reasonable to identify it as the CEP of the chiral phase transition (solid trajectory) below the given $L$. The second inflection point is induced by the first mode when entering below the Fermi surface. For sufficiently small $L$, this new CEP also turns to lower $T$ and higher $\mu_q$ (like LSM A and LSM B), while the original CEP is continued toward the $\mu_q$ axis generated by the second lowest modes when entering below the Fermi surface. Similar nonmonotonic behavior was also found in \cite{Almasi:2016zqf} within a functional renormalization group approach. For LSM A and LSM B, the original CEPs start from a much lower $\mu_q$ and higher $T$, so their trajectories continuously connect the CEP at $L=\infty$ to the CEP with the smallest possible $L$. The quasi-CEPs (second-order points at lower $\phi_N$ values) of the transitions generated by the second, third, etc. lowest modes (when entering below the Fermi surface) are also present for LSM A and B, but their presence remains hidden. The $1/L$ dependence of $\mu_q^\text{CEP}$ and $T^\text{CEP}$ is also shown separately for each model in Fig.~\ref{fig:TmuCEPs}.
It can be seen that for $1/L<0.2~\text{fm}^{-1}$ ($L>5$~fm) the curves are flat. For $0.2~\text{fm}^{-1}<1/L<0.5~\text{fm}^{-1}$ ($2~\text{fm}<L<5$~fm) for the ePQM and LSM C models, the original CEP disappears, while a newly generated CEP emerges and becomes dominant. In this range the $T^\text{CEP}$ ($\mu_q^\text{CEP}$) is continuously decreasing (increasing) for LSM A and B. Finally, for $1/L>0.5\,\text{fm}^{-1}$ ($L<2$~fm) both $\mu_q^\text{CEP}$ and $T^\text{CEP}$ show a very similar trend for each model. This also shows that for sufficiently small sizes the CEP corresponds to the second-order point belonging to the highest $\phi_N$ (first step in Fig.~\ref{fig:sigma_in_mu_Dp200}) of the transition, which is generated by the first mode when entering below the Fermi surface.

%%%%%%%%%%%%%%%%%%%%%%%%%%%%%%%%%%%%%%%%%%%%%%%%%%%%%%%%newSubSubSec
\subsubsection{Size dependence of the CEP with PBC}

\begin{figure}[htbp]
    \centering
     \includegraphics[width=0.48\textwidth]{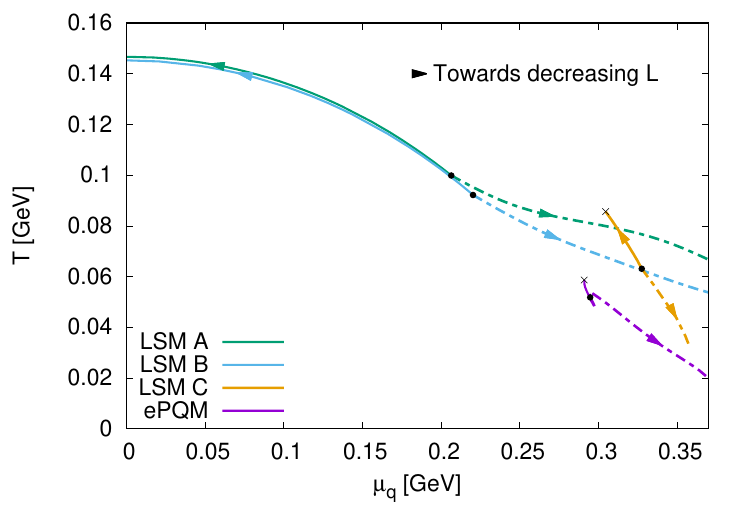}
    \caption{The path of the critical endpoints over the phase diagram for the different models with PBC. Arrows point to lower values of $L$}
    \label{fig:CEPs_path_PBC}
\end{figure}

For PBC with a vacuum term of infinite size, the CEP moves to higher temperatures and lower quark chemical potentials with decreasing system size, leading to an expansion of the first-order region for each model, as shown (with solid lines) in Fig.~\ref{fig:CEPs_path_PBC}. 
However, the appearance of the new solution of the field equations discussed in Sec.~\ref{sSec:Discretization_finite_temp_and_chempot} turns the phase transition, either crossover or first order, into an unphysical first-order transition in each model at roughly the same size around $L=5.5$ fm (at $L=5.46$ for the ePQM), thus we cannot continue the calculation to smaller sizes. 
For LSM A and B models, the CEP reaches the $T$ axis, leaving a first-order phase transition in the full phase boundary well before the nonphysical effect becomes dominant. For the LSM C and the ePQM models, there is still a crossover region where this unphysical transition occurs and the CEP is still very far from the $T$ axis. This CEP is marked with a black cross in the figure. 

As mentioned in Sec.~\ref{sSec:Discretization_finite_temp_and_chempot}, by including the effect of the discretization in the vacuum part only in the zero mode, the nonphysical transition disappears. Furthermore, this zero-mode correction of the vacuum contribution\footnote{Which is a discretization of the vacuum part only for the first mode for ePQM, but an extra contribution for LSM A, B, and C, since these models have no vacuum part at all.} is large enough to even reverse the trend of the CEP trajectory as $L$ is decreased. Therefore, it moves (for ePQM after some transient oscillations) to higher temperatures and lower chemical potentials with decreasing size, resulting in the reduction of the first-order region. The trajectory of the CEPs in this scenario is also shown in Fig.~\ref{fig:CEPs_path_PBC} with the dash-dotted lines. For LSM A, this approximation reproduces the results in \cite{Palhares:2009tf} for PBC. We note that for low $T$ and high $\mu_q$ there are multiple solutions of the FEs [Eqs.~\eqref{Eq:FE_phiN} and \eqref{Eq:FE_phiS}] due to the structure generated by the modes entering below the Fermi surface, which was discussed in the previous section (see Fig~\ref{fig:sigma_in_mu_Dp200}). However, in this case, the structure of $\phi_{N/S}(T,\mu_q)$---and hence the structure of the order parameter $\Delta(T,\mu_q)$---is more complicated than in the case of APBC with infinite vacuum, and it is difficult to identify well-defined phase transitions for low $T$ and high $\mu_q$. Therefore, when the $T=0$ behavior begins to affect the $T=T^\text{CEP}$ region at sufficiently small sizes---below $L=$4-5 fm for LSM C and the ePQM models, and below $L=2$ fm for the LSM A and B models---the CEP will not be tracked further with decreasing size.

For PBC-0, the influence of the missing zero mode of the fermionic thermal contribution---implying a cubic low-momentum cutoff---will be dominant. It overwhelms any effect of summation and pushes the CEP toward the chemical potential axis. However, the appearance of the new CEP arising from the $T=0$ behavior influences the path of the critical endpoint also for this boundary condition. The location of the CEP for very small system sizes is determined only by the first mode---following the absent zero mode---entering below the Fermi surface.

%%%%%%%%%%%%%%%%%%%%%%%%%%%%%%%%%%%%%%%%%%%%%%%%%%%%%%%%NEWSEC
\section{Conclusion}
\label{Sec:Conclusion}

In this paper we have investigated the effects of finite system size in the framework of the vector and axial vector meson extended Polyakov linear sigma model by using two types of implementations: a low-momentum cutoff of the momentum integrals and a discretization of the momentum space implied by the finite spatial extent of the system. The effect of these modifications on the phase diagram of strongly interacting matter, the role of the vacuum term and the different boundary conditions have been studied in detail. 

First, the case of low-momentum cutoff was investigated. In particular, the effect of the size of the vacuum, either finite or left unchanged (infinite size), was studied. The latter leads to similar results as in previous studies where the fermionic vacuum corrections were not considered. It was found that the modification of the vacuum integrals strongly pushes the system toward the chirally symmetric phase. Therefore, the whole phase boundary moves to lower temperatures until the chirally broken phase vanishes around $L\approx 2$ fm, where a crossoverlike transition can be observed in the vacuum physical quantities. However, if the vacuum is kept in its infinite form, the chirally broken phase even extends as the boundary moves to higher $T$ and $\mu_q$. In both cases, the critical endpoint moves to lower temperatures and higher chemical potentials. For a vacuum term of finite size, it turns toward the chemical potential axis and even disappears around $L\approx 2.5$ fm. On the other hand, for an infinite size vacuum, the CEP turns toward high $\mu_q$, and although the first-order line shrinks with decreasing size, the CEP does not reach the axis until at least $L\approx 0.5$ fm. The volume dependence of the phase diagram and the order of the chiral phase transition were also studied in and near the chiral limit, i.e., the $m_{u,d}=0,~m_s\neq 0$ axis of the Columbia plot. It was found that the phase structure changes with decreasing size in a very similar way to the physical case, while a second-order phase transition is found only at vanishing explicit breaking and crossover elsewhere.

We also studied the modification of the baryon fluctuations at finite volume in the low-momentum cutoff scenario. At $\mu_q=0$ we found a smoothening of the kurtosis as a function of temperature, which was expected. The kurtosis and the skewness were also calculated through the critical endpoint ($\mu_q = \mu_q^{\text{CEP}}$ or $T=T^{\text{CEP}}$) as a function of $T/T^{\text{CEP}}$ or $\mu_q/\mu_q^{\text{CEP}}$. $\kappa (T/T^\text{CEP})\big|_{\mu_q=\mu_q^\text{CEP}}$ showed an increasing behavior on both sides, not very close to the CEP. However, we concluded that this trend might be a consequence of the rescaling by $T^\text{CEP}$, since for small volumes the critical endpoint moves to lower $T$ values. For $\kappa (\mu_q/\mu_q^{\text{CEP}})\big|_{T=T^\text{CEP}}$ the rescaling by $\mu_q^{\text{CEP}}$ did not alter the results and a slight decrease was observed near the critical endpoint, which might be also connected to the significant shift of the CEP. Therefore, we claimed that the kurtosis probably does not possess a strong size dependence, which is consistent with what has been found in the literature using a DS approach.

As an alternative to the low-momentum cutoff, the discretization of the momentum space was also studied. On the one hand, it turned out that the implementation of this approach is more complicated, and on the other hand, in several cases, the solution of the field equations was lost. Moreover, the $\mu_q$ dependence of the condensates showed a more complicated, staircaselike structure than in the low-momentum cutoff case, due to the different modes falling below the Fermi surface---which may be an artifact of our assumption due to the size-independent Fermi surface, but could also be realized in certain physical systems, e.g., in solid state physics. This behavior makes it more difficult to define a unique transition point and causes some unexpected behavior in the trajectory of the CEP as a function of the characteristic size $L$, and the resulting finite size effects are not conclusive. The choice of different boundary conditions, such as PBC, APBC, and PBC-0, also causes significant changes in the trajectory of the CEP. We have also compared the behavior of our model with other models in the literature. 

Comparing the predictions for the size dependence of the CEP for the low-momentum cutoff with and without modification of the vacuum part (see Fig.~\ref{fig:phase_diag_cutoff_both}) and for the discretization with APBC (Fig.~\ref{fig:CEPs_path_APBC}) and PBC (Fig.~\ref{fig:CEPs_path_PBC}), it can be seen that completely different behavior can be found. The role of the treatment of the vacuum term and the different boundary conditions is paramount in the exact trajectory of the CEP as $L$ decreases. The only similarity seems to be that for small $L$ the CEP moves toward the $\mu_q$ axis, as expected.

Since the present results do not provide a consistent picture of the details of the finite size effects on the phase diagram, we need to find better approaches to this problem. For this purpose, in addition to improving the models used, we need more sophisticated methods to implement these effects, e.g., by considering more physically motivated shape and boundary conditions for the system. Furthermore, besides the momentum space constraints, one could consider the finite size dependence in direct space as well. However, this would lead to a nonhomogeneous potential and hence to nonhomogeneous meson condensates, which would require overcoming new obstacles.

%%%%%%%%%%%%%%%%%%%%%%%%%%%%%%%%%%%%%%%%%%%%%%%%%%%%%%%%NEWSEC
\section{Acknowledgment}

We would like to thank K. Fukushima and L. Turko for valuable discussions. The research was supported by the Hungarian National Research, Development and Innovation Fund under Project No. FK 131982 and under Project No. K 138277. K.R. and P.M.L. acknowledge support from the Polish National Science Center (NCN) under Opus grant no. 2022/45/B/ST2/01527. G. K. thanks the PHAROS COST Action (CA16214) for partial support. We also acknowledge support for the computational resources provided by the Wigner Scientific Computational Laboratory (WSCLAB).

%%%%%%%%%%%%%%%%%%%%%%%%%%%%%%%%%%%%%%%%%%%%%%%%%%%%%%%%Appendices
\appendix

%%%%%%%%%%%%%%%%%%%%%%%%%%%%%%%%%%%%%%%%%%%%%%%%%%%%%%%%NEWSEC
\section{Order of the transition in the chiral limit for finite size} \label{Sec:chiral_limit}

It is expected that the chiral phase transition, which is a crossover for the explicitly broken theory, at small chemical potential turns to a second-order phase transition as the explicit breaking is removed. As seen from the mean-field level Landau theory of phase transitions, this happens only in the explicitly restored limit. Since in our case also a mean-field approximation is used, it is not surprising that we see a second-order phase transition only in the chiral limit.

However, in recent works \cite{Feng:2019juf, Hu:2021upi} using a different model the authors found that for finite sizes the transition becomes second-order already for a small but nonzero explicit breaking. Accordingly, we investigate the order of the chiral phase transition in and close to the chiral limit using the low-momentum cutoff scenario to mimic the impact of finite system size.

The explicit breaking of the chiral symmetry in the ePQM model is governed by the external fields $h_{N/S}$ which are coupled to a single meson field in the Lagrangian \eqref{eq:ELSM_Lag}. Thus, to study the removal of the explicit breaking and the chiral limit one has to start with the physical value of $h_N$ determined by the parametrization and reduce this parameter until reaching $h_N=0$. The field equation for $\phi_N$ in Eq.~\eqref{Eq:FE_phiN} can be rewritten in the form
\be \label{eq:FE_phiN_alt}
m_\pi^2 \frac{\phi_N}{Z_\pi^2} - h_N =0\text{,}
\ee 
where $Z_\pi$ is the pion wave function renormalization constant that is
connected with the pion decay constant by $f_\pi = \phi_N/Z_\pi$.
For $h_N\rightarrow 0$, and nonzero spontaneous symmetry breaking (i.e., $\phi_N>0$), the pion mass has to vanish as expected in the chiral limit (Goldstone theorem). We note that to connect the parameter $h_N$ to the current quark mass, which does not appear naturally in a quark-meson model, a possible way is provided by the Gell-Mann--Oakes--Renner (GMOR) relation, $m_\pi^2 f_\pi^2 = -2m_0 \langle \bar q q \rangle$, where $m_0$ is the current quark mass, while $ \langle \bar q q \rangle$ is the chiral condensate. According to Eq.~\eqref{eq:FE_phiN_alt} the GMOR relation can be rewritten as $h_N \phi_N = -2m_0 \langle \bar q q \rangle$, which directly connects the product of the explicit and spontaneous symmetry breaking parameters in the linear sigma model to the parameters in a quark-based model, like the NJL. Although $h_N\phi_N$ gives the expected physical value for the product, to obtain $m_0$  ($\equiv m_u$) the identification of the $\langle \bar q q \rangle$ chiral condensate would be needed in the quark-meson model separately. However, in the ePQM model, the quark-antiquark condensate is included in the meson masses via the fermion one-loop correction. Therefore, the parametrization that is based on the meson masses fixes the value of the chiral condensate, which turns out to be too low in our case.

\begin{figure}[ht]
    \centering
     \includegraphics[width=0.48\textwidth]{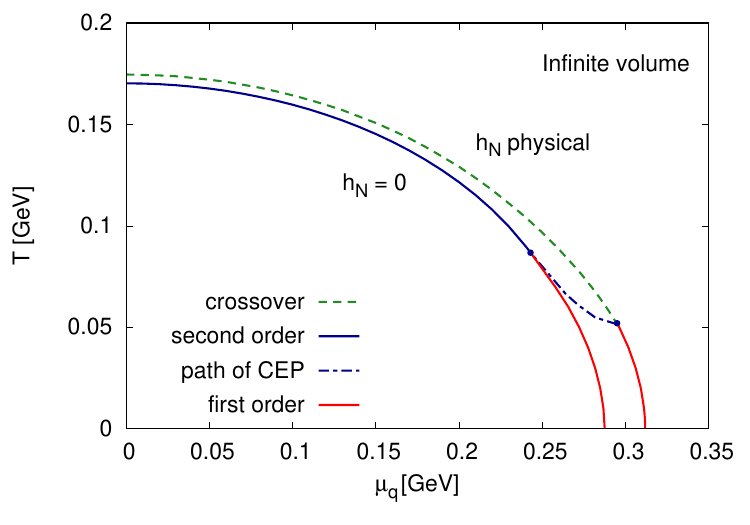}
    \caption{The phase diagram in the physical case and in the chiral limit, which is given by $h_N=0$ in the ePQM model.}
    \label{fig:phase_diag_hN}
\end{figure}

\begin{figure}[tbp]
    \centering
     \includegraphics[width=0.48\textwidth]{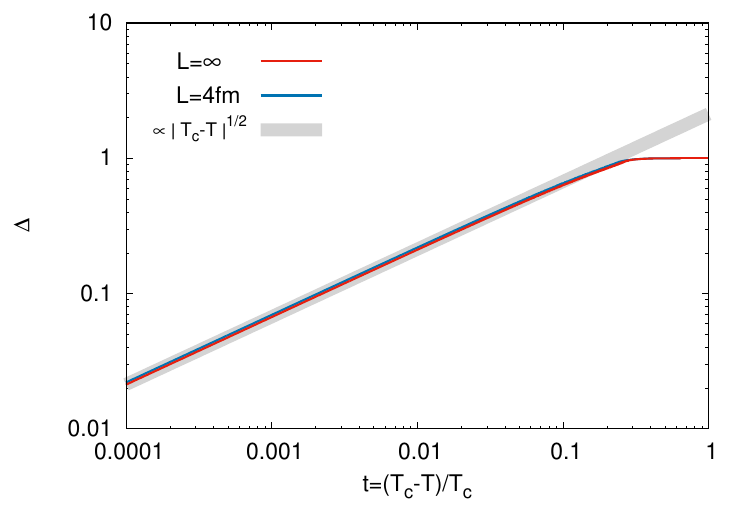}
     \caption{The critical scaling of the order parameter for the chiral symmetry in the second-order phase transition at $h_N=0$ for infinite and $L=4$~fm sizes. The gray band shows scaling for the mean-field theory with critical exponent $\beta=1/2$. }
    \label{fig:scaling_of_Delta}
\end{figure}

\begin{figure}[htbp]
    \centering
     \includegraphics[width=0.48\textwidth]{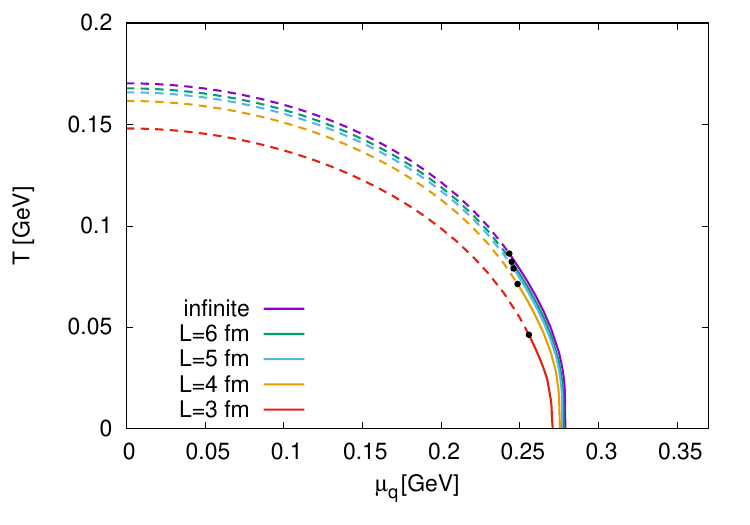}
    \caption{The phase diagram for several finite sizes in the chiral limit. Here the dashed lines correspond to second-order phase boundary}  
    \label{fig:phase_diag_at_hN0_L}
\end{figure}

In Fig.~\ref{fig:phase_diag_hN}, the chiral phase boundary is shown in the chiral limit for the $N_f=2\text{ (light)}+1\text{ (heavy)}$ case, i.e., for $h_N\to 0$, $h_S=h_S^\text{phys}$ (left axis on the Columbia plot). 
It can be seen that the transition at vanishing $\mu_q$ moves to lower temperature with the decreasing $h_N$ and the crossover becomes a second-order phase transition at $h_N=0$, which is consistent with the behavior found in previous studies \cite{Halasz:1998qr, Hatta:2002sj, Philipsen:2021qji}. 
The CEP moves to lower quark chemical potentials and higher temperatures and ends up in a tricritical point, as it is expected. The universality class of the second-order phase transition is of the mean-field theory, which can be seen in Fig.~\ref{fig:scaling_of_Delta}, where the scaling of the order parameter $\Delta$ (subtracted condensate) is presented in the chiral limit. 
In Fig.~\ref{fig:phase_diag_at_hN0_L}, the volume dependence of the phase boundary can be seen in the chiral limit. The solid and dashed lines correspond to first- and second-order phase transitions, respectively, while the black dots now mark the tricritical points. It is clear that the qualitative picture is the same as in the case of physical explicit breaking shown in the top panel of Fig.~\ref{fig:phase_diag_cutoff_both}.

\begin{figure}[tbp]
    \centering
     \includegraphics[width=0.48\textwidth]{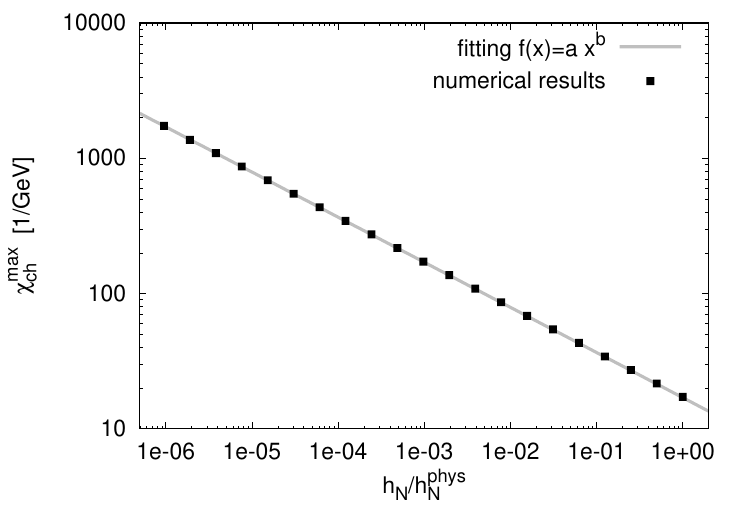}
     \caption{The maximum of the chiral susceptibility as a function of the explicit breaking $h_N/h_N^\text{phys}$}
    \label{fig:chir_susc_max}
\end{figure}

A possible method to find the order of a phase transition is to look at the chiral susceptibility, i.e., the derivative of the order parameter with respect to the temperature $\chi_\text{ch}=\partial \Delta /\partial T$. $\chi_\text{ch}$ diverges in the case of a second-order phase transition at the transition point. Although in a numerical calculation the divergence can never be seen explicitly, for any $h_N>0$ it can be shown that the maximum of $\chi_\text{ch}$ is finite for any investigated $L$ values. The $h_N/h_N^\text{phys}$ dependence of the maximum of the chiral susceptibility is shown in Fig.~\ref{fig:chir_susc_max} for $L=4$ fm. 
It seems from the log-log plot that there will be no divergence for finite $h_N$. The points can be fitted with the form $a(x+c)^b$ for $x=h_N/h_N^\text{phys}$. However, the $c$ parameter is getting smaller and smaller when the number of points is increased or the accuracy of the computation is improved. The best fit can be obtained by setting $c=0$, in which case $a=16.99$ and $b=-0.33$ is found. Furthermore, from the field equation in Eq.~\eqref{Eq:FE_phiN} it is clear that for any $T<\infty$ temperature and for a nonvanishing explicit breaking $\phi_N=0$ cannot be a solution. Therefore, the order parameter will not vanish for finite temperatures, resulting in an increasingly sharper but still crossover transition when $h_N$ is decreased. On the other hand, we also emphasize that at $h_N=0$ one finds a second-order phase transition even for finite volumes.

%%%%%%%%%%%%%%%%%%%%%%%%%%%%%%%%%%%%%%%%%%%%%%%%%%%%%%%%NEWSEC
\section{Solid angle integration from a cube to a sphere}
\label{Sec:app_solid_angle}

\begin{figure}[htbp]
    \centering
    \includegraphics[width=.15\textwidth]{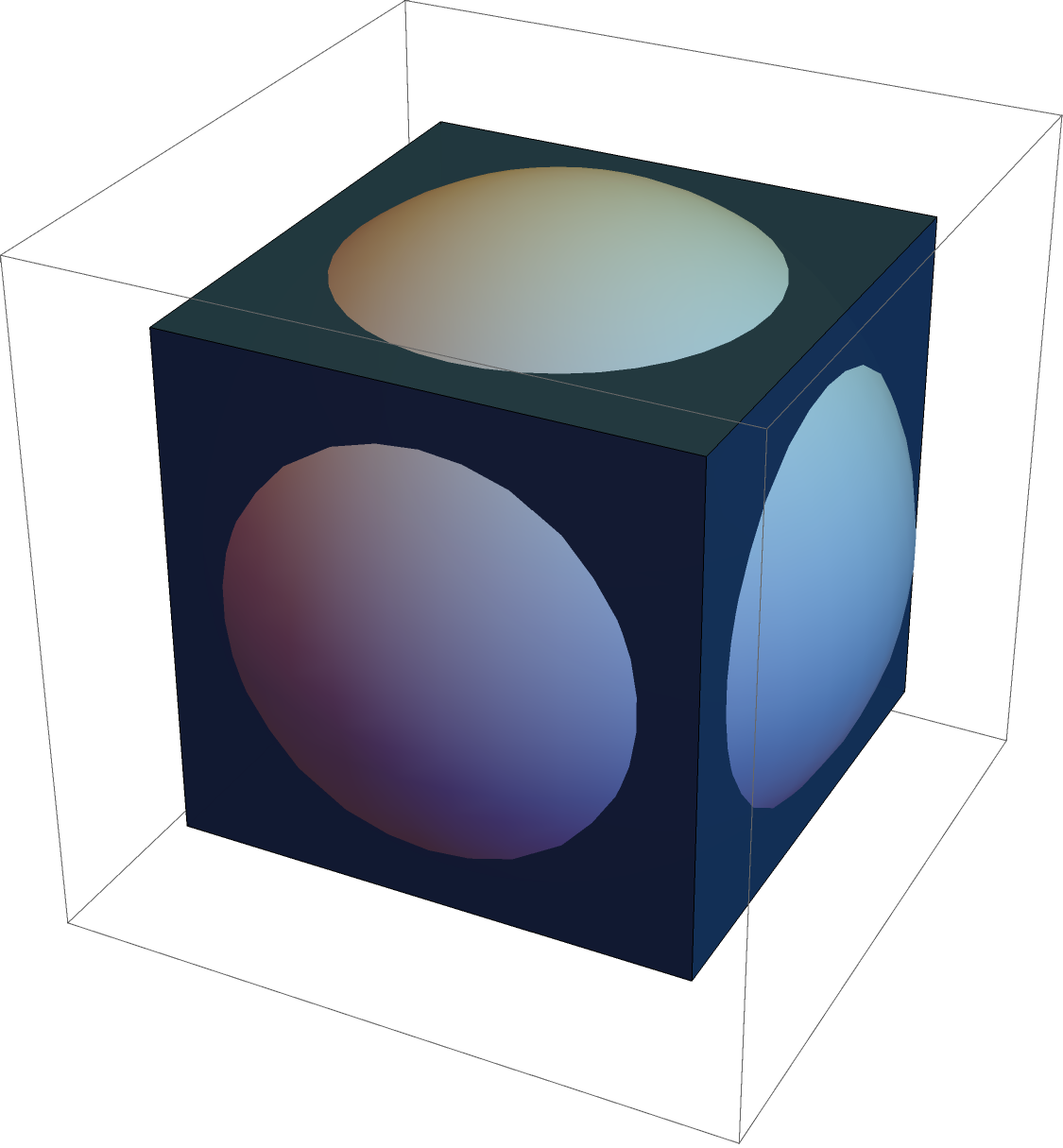}
    \includegraphics[width=.15\textwidth]{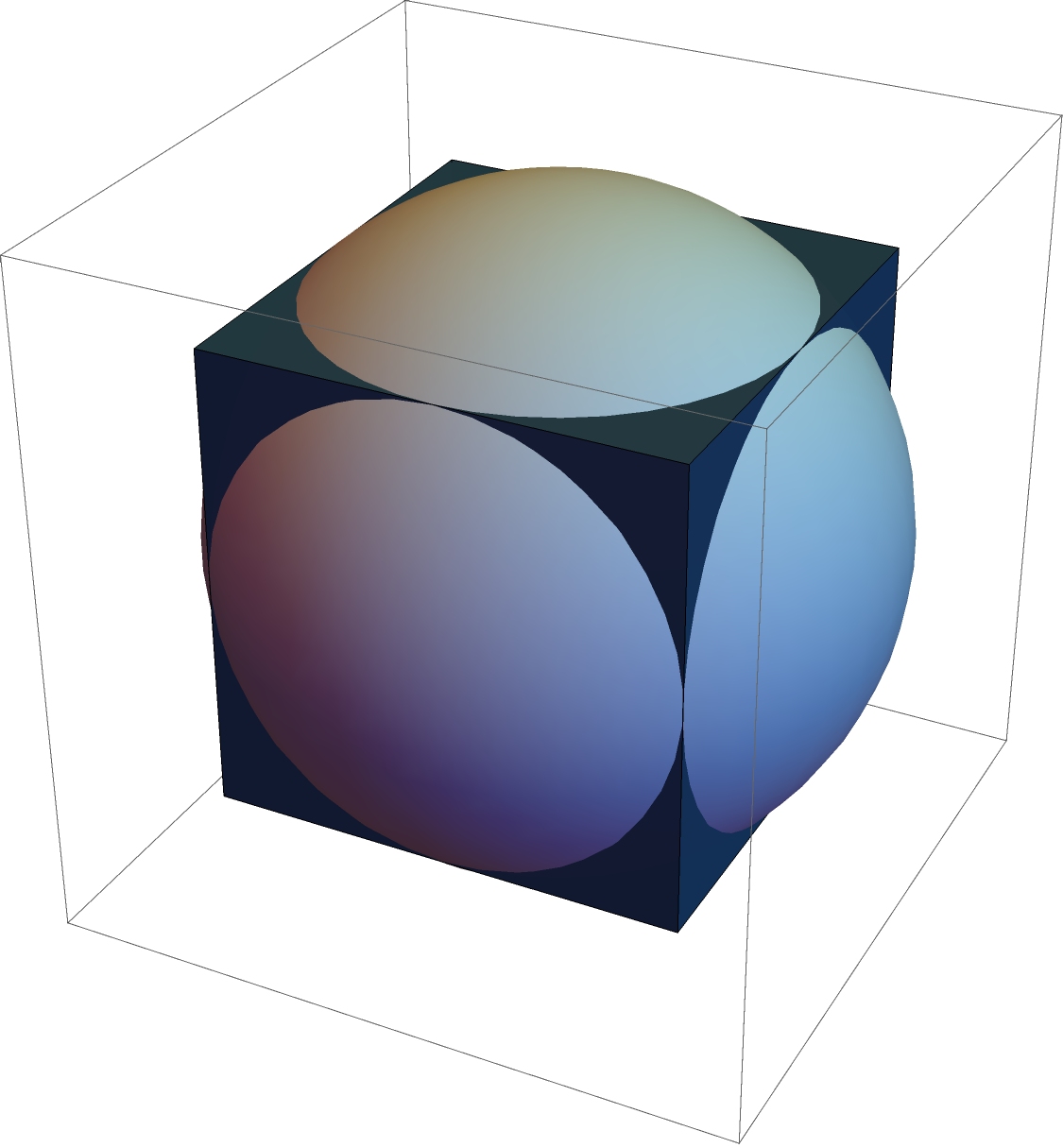}
    \includegraphics[width=.15\textwidth]{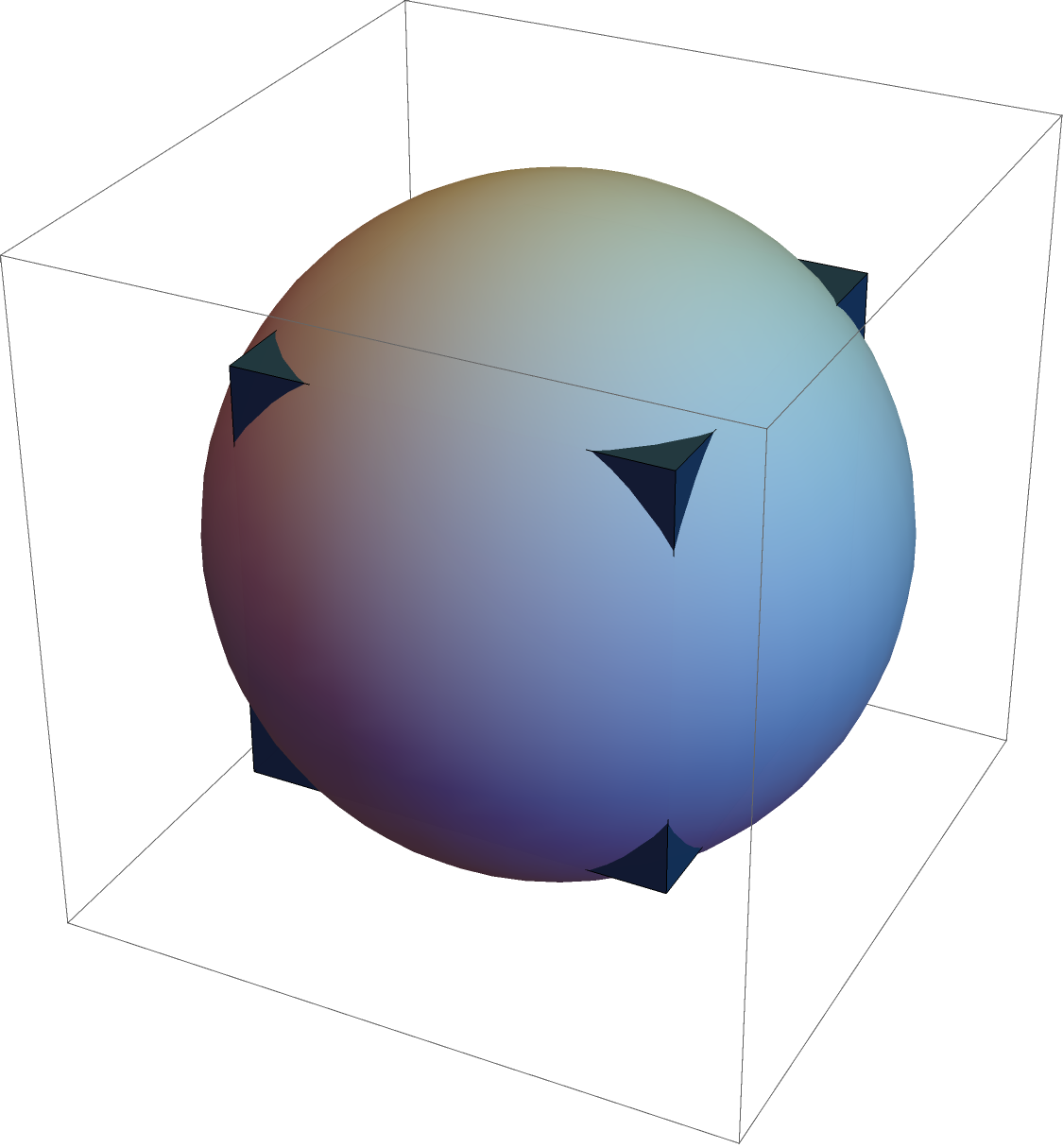}
    \caption{Intersection of a cube with side length $2\lambda_{\Sigma}$  and a sphere with radius $R=p$ for  $\lambda_{\Sigma}<p<\sqrt{2}\lambda_{\Sigma}$ (left), $p=\sqrt{2}\lambda_{\Sigma}$ (middle), and $\sqrt{2}\lambda_{\Sigma} <p< \sqrt{3} \lambda_{\Sigma}$ (right).}
    \label{Fig:Solid_angles}
\end{figure}

In this section, Eqs.~\eqref{Eq:solid_angle_int}--\eqref{Eq:solid_angle_int_p2} are derived, which are the result of a solid angle integral. Imagine a cube with side length $2\lambda_{\Sigma}$ and draw a sphere with a radius $R=p$ which is in the interval $\lambda_{\Sigma}<p<\sqrt{3}\lambda_{\Sigma}$. If $p=\lambda_{\Sigma}$, the sphere is inside the cube and touches its sides, while if $p=\sqrt{3}\lambda_{\Sigma}$, the sphere includes the cube and touches its vertices. In between, part of the sphere is outside the cube, while part is inside. Our goal is to calculate -- for a general $p$ in the interval mentioned -- the area of the sphere that is outside of the cube. If we divide the resulting area by $p^2$, we get the desired solid angle $\Omega=S/p^2$. This is illustrated in Fig.~\ref{Fig:Solid_angles}. 

\begin{figure}[htbp]
    \centering
    \includegraphics[width=.2\textwidth]{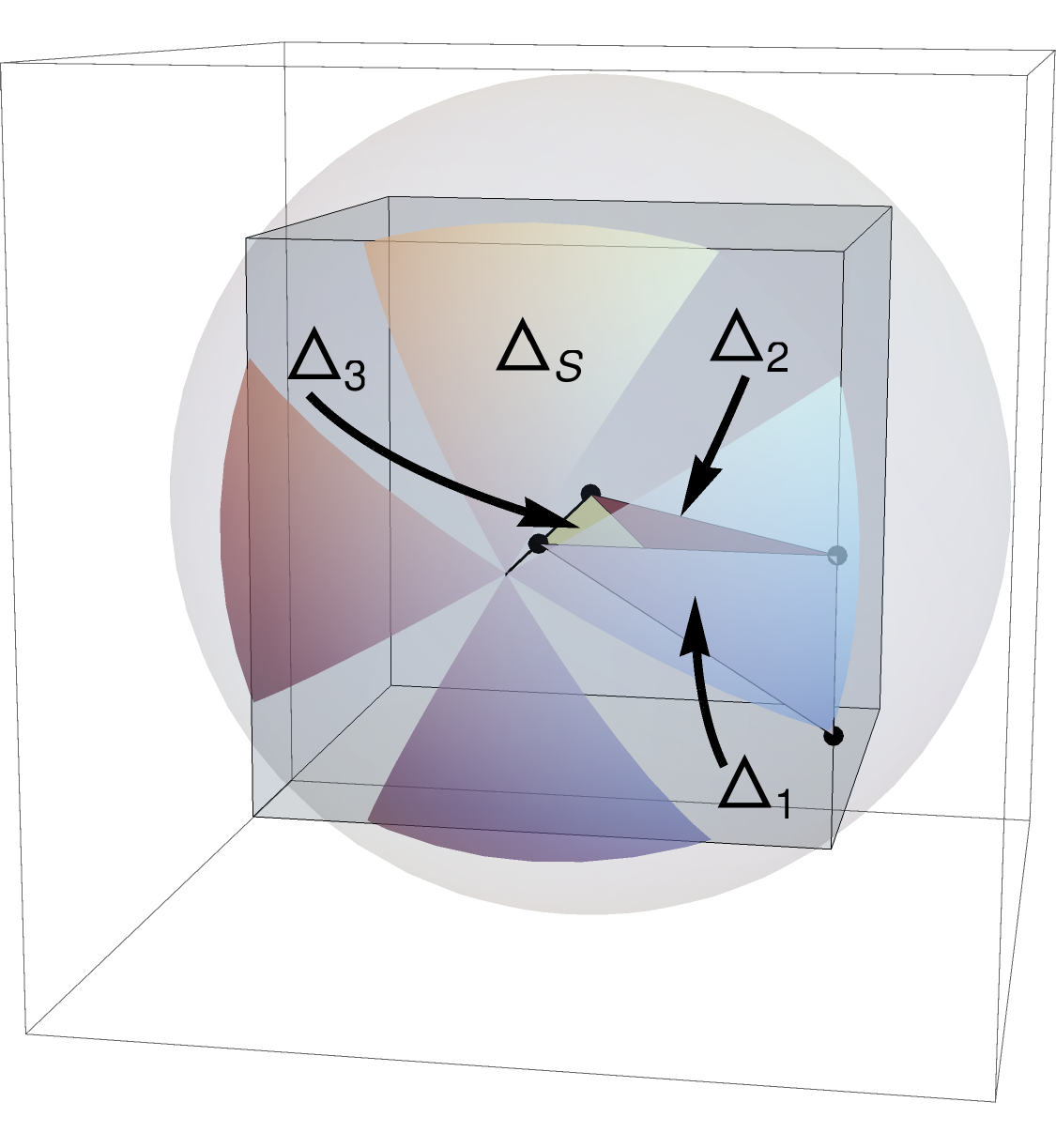}
    \includegraphics[width=.2\textwidth]{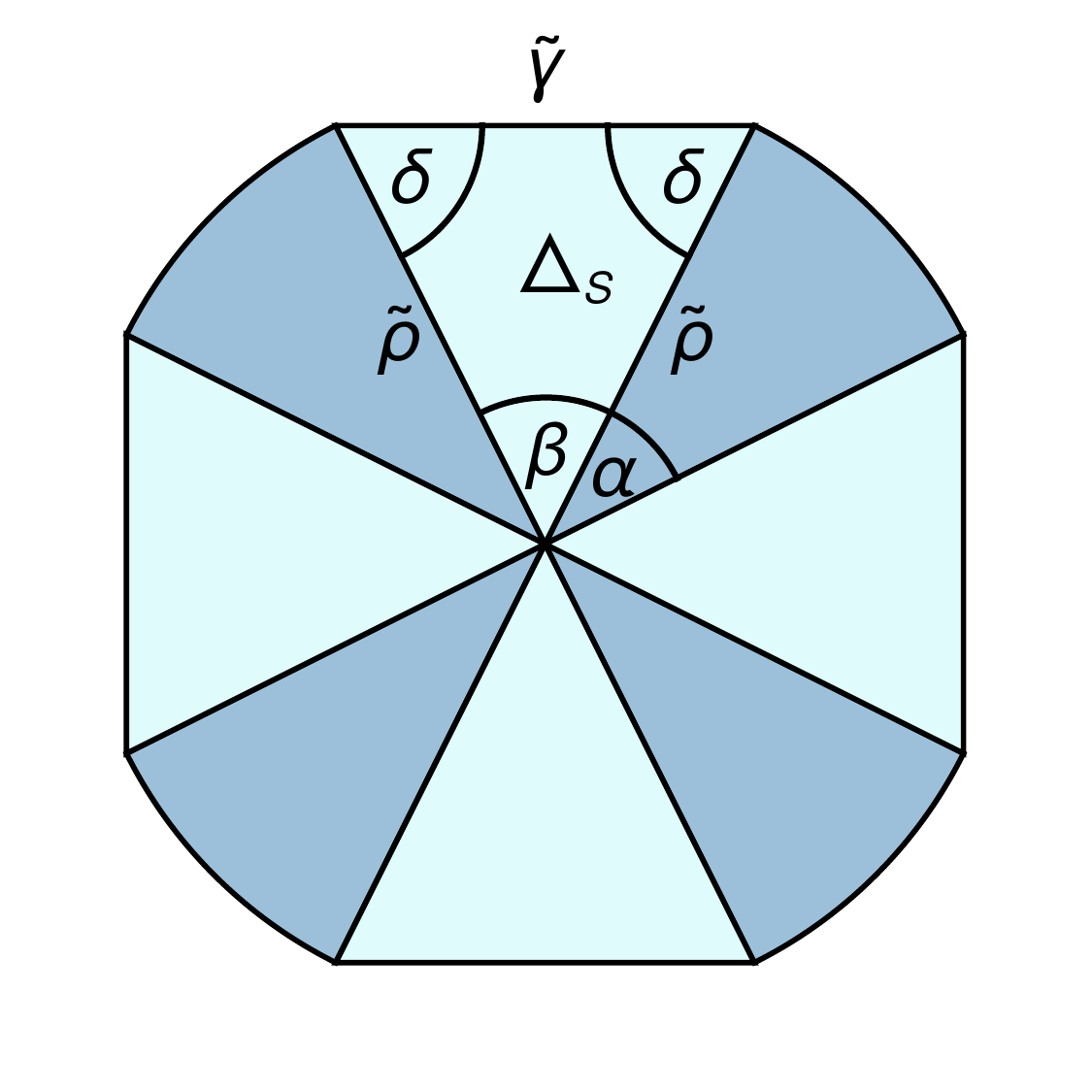}
    \includegraphics[width=.15\textwidth]{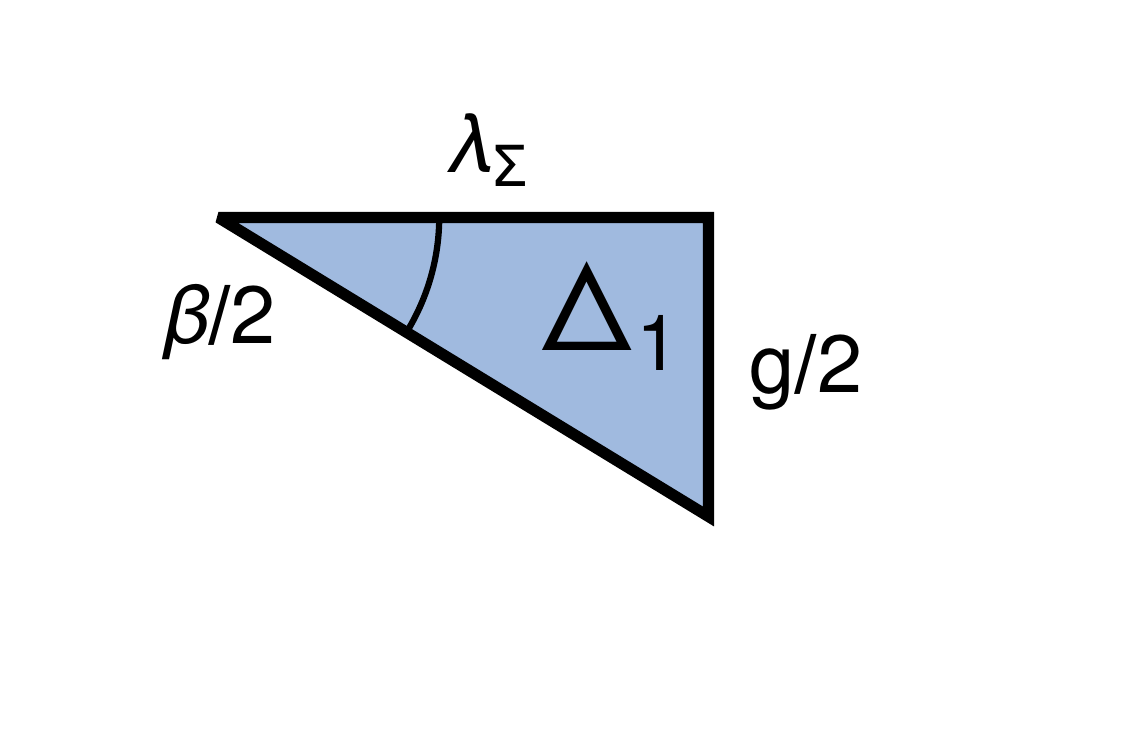}
    \includegraphics[width=.15\textwidth]{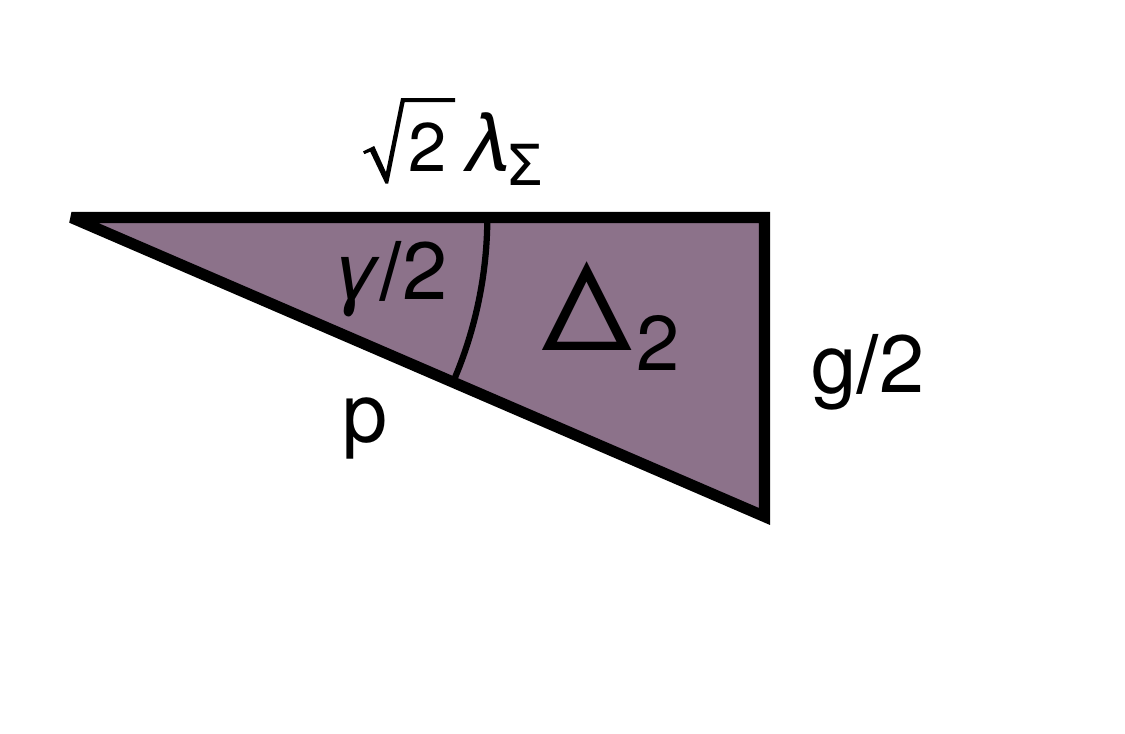}
    \includegraphics[width=.15\textwidth]{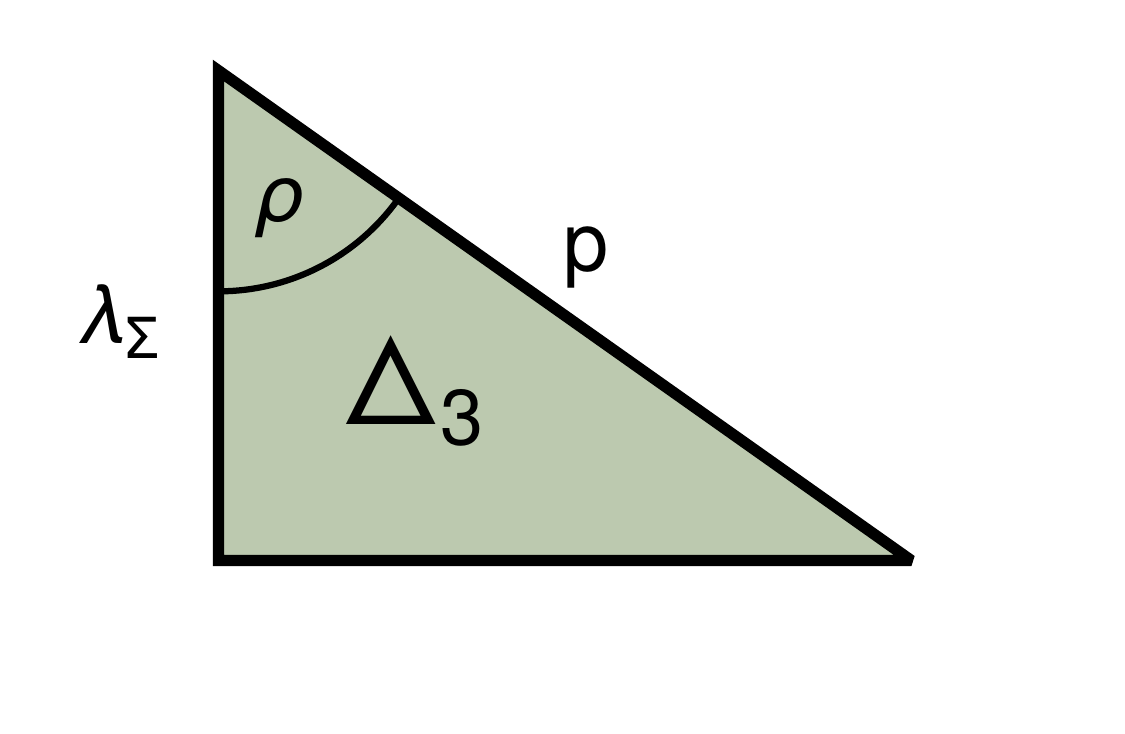}
    \caption{Division of the modified caps into spherical triangles and sectors (top) and the right-angled triangles (bottom).}
    \label{Fig:Reduced_cap}
\end{figure}

For $\lambda_{\Sigma}<p\leq\sqrt{2}\lambda_{\Sigma}$ there is a cap on each side of the cube. The surface of this spherical zone is $S_\text{zone}=2\pi Rh$ with height $h=p-\lambda_{\Sigma}$ and $R=p$, from which the full solid angle in this $p$ range is given by
\be 
\label{Eq:surface_of_a_cap}
\Omega^{\lambda_{\Sigma}}_1 (p)=6 S_\text{zone}/p^2=12 \frac{\pi (p-\lambda_{\Sigma})}{p} \text{ .}
\ee 
At $p=\sqrt{2}\lambda_{\Sigma}$ the caps touch each other. As the momentum increases, they overlap, so either the double counting must be removed or another method is needed. In both cases, one has to calculate spherical triangles and circular sectors of a spherical cap. We will calculate the contribution of modified caps that avoid double counting by dividing them into four congruent spherical triangles $\triangle_S$ (shown as ordinary triangles when drawn in a plane for simplicity) and four congruent sectors, as can be seen in Fig.~\ref{Fig:Reduced_cap}. To determine the parameters of $\triangle_S$ and the angle of the circular sectors, one can use the right-angled triangles $\triangle_{1-3}$ also shown in the same figure. 
The solid angle covered by a spherical triangle is equal to its spherical excess, which in our case is $E_{\triangle_S}=2\delta+\beta -\pi$. The angle $\beta$ can be calculated from the right triangle $\triangle_1$ on the side of the cube with vertices at the center of the side, at half of the edge of the cube, and at the point on the edge where the sphere intersects it. Its unknown leg with length $g/2$ can be obtained from the right triangle $\triangle_2$, which has the same vertices on the edge of the cube, but the third one is in the center of the cube. With a straightforward calculation, one can get
\be 
\label{Eq:app_beta_def}
\beta = 2\arctan \sqrt{p^2/ \lambda_\Sigma^2-2}\text{ .}
\ee 
We obtain $\delta$ using Napier's rule for right spherical triangles \cite{Todhunter:1886},
\be 
\tan(\gamma/2) =\cos(\delta)\tan(\rho),
\ee 
where $\gamma$ and $\rho$ are the angles of the arcs, $\tilde\gamma=p\gamma$ and $\tilde\rho=p\rho$, which are the sides of $\triangle_S$. They can be expressed with $p$ and $\lambda_{\Sigma}$, with using $\triangle_2$ and $\triangle_3$ (with vertices at the center of the cube, at the side of the cube, and at the intersection of the sphere and the edge of the cube), respectively, to have
\begin{align}
    \gamma=&2 \arccos\left(\sqrt{2}\lambda_{\Sigma}/p\right)\text{ ,}\\
    \rho=&\arccos\left( \lambda_{\Sigma}/p\right)\text{ ,}
\end{align}
which results in
\be 
\delta= \arccos \sqrt{\frac{p^2-2 \lambda_\Sigma^2}{2p^2-2\lambda_\Sigma^2}}\text{ .}
\ee 
\vfill\null
Thus, the excess of the spherical triangle becomes 
\begin{align}
E_{\triangle_S}= &2 \arccos \sqrt{\frac{p^2-2 \lambda_\Sigma^2}{2p^2-2 \lambda_\Sigma^2}} \\
&+ 2\arctan \sqrt{p^2/ \lambda_\Sigma^2-2} -\pi \text{ .}
\end{align}

The contribution of a circular sector of a spherical cap can be given by the solid angle of the cap itself normalized by the ratio of the inner angle $\alpha$ and $2\pi$. Using $4\alpha =2\pi-4\beta$, the four sectors of a cap give 
\be 
\Omega_{4\sector}=\left(2\pi-8\arctan \sqrt{p^2/ \lambda_\Sigma^2-2} \right) (p-\lambda_{\Sigma})/p\text{ ,}
\ee 
and thus the full solid angle of the sphere outside the cube for $\sqrt{2} \lambda_{\Sigma} < p < \sqrt{3} \lambda_{\sigma}$ is finally given by  
\begin{align}
    \Omega^{\lambda_{\Sigma}}_2(p)=
    \frac{4}{p}\Bigg( &12p\arccos{\sqrt{\frac{p^2-2 \lambda_\Sigma^2}{2p^2-2 \lambda_\Sigma^2}}}
    \nonumber \\
    &+12\lambda_{\Sigma}\arctan\sqrt{p^2/\lambda_{\Sigma}^2-2} \nonumber \\
    &-3\pi(\lambda_{\Sigma}-p)\Bigg)\text{ .}
\end{align} 

\FloatBarrier

\bibliography{FinV_ELSM}

\end{document}